\documentclass[10pt, two column, twoside]{IEEEtran}
\usepackage{makecell}
\usepackage{lipsum}
\usepackage{cuted}
\usepackage{amsthm}
\usepackage{amsmath}
\usepackage{algorithmic}
\usepackage{algorithm}
\usepackage{mathrsfs}
\usepackage{graphicx}
\usepackage{subfigure}
\usepackage{cite}
\usepackage{bm,epsfig,amsthm,url}
\usepackage{indentfirst}
\usepackage{amssymb}
\usepackage{epstopdf}
\usepackage{stfloats}  
\usepackage{xcolor}
\usepackage{mathtools}
\usepackage{hyperref}
\usepackage{verbatim}
\usepackage{enumerate} 
\usepackage{caption}

\usepackage{makecell}

\newtheorem{theorem}{Theorem}

\DeclareMathOperator{\diag}{diag}

\definecolor{MyColor}{RGB}{0, 0 , 255} 
\definecolor{MyColor2}{RGB}{0, 0 , 0} 
\definecolor{MyColor3}{RGB}{0, 255 , 0} 

\usepackage{changes} 

\allowdisplaybreaks

\begin{document}

\title{Extremely Large-Scale Movable Antenna-Enabled Multiuser Communications: Modeling and  Optimization}
\author{Min~Fu,~\IEEEmembership{Member,~IEEE}, Lipeng~Zhu,~\IEEEmembership{Member,~IEEE}, and~Rui~Zhang,~\IEEEmembership{Fellow,~IEEE}
	\thanks{M. Fu and L.~Zhu are with the Department of Electrical and Computer Engineering, National University of Singapore, Singapore 117583 (e-mails:  fumin@nus.edu.sg and zhulp@nus.edu.sg).}
			\thanks{R. Zhang is with School of Science and Engineering, Shenzhen Research Institute of Big Data, The Chinese University of Hong Kong, Shenzhen, Guangdong 518172, China (e-mail: rzhang@cuhk.edu.cn). He is also with the Department of Electrical and Computer Engineering, National University of Singapore, Singapore 117583 (e-mail: elezhang@nus.edu.sg).
}
}

\maketitle

\setlength\abovedisplayskip{2pt}
\setlength\belowdisplayskip{2pt}
\setlength\abovedisplayshortskip{2pt}
\setlength\belowdisplayshortskip{2pt}
\setlength\arraycolsep{2pt}

\begin{abstract}
Movable antenna (MA) has been recognized as a promising technology to improve communication performance in future wireless networks such as 6G. To unleash its potential, this paper proposes a novel architecture, namely extremely large-scale MA (XL-MA), which allows flexible antenna/subarray positioning over an extremely large  spatial region for effectively enhancing near-field effects and spatial multiplexing performance.
In particular, this paper studies an uplink XL-MA-enabled multiuser system, where single-antenna users distributed in a coverage area are served by a base station (BS) equipped with multiple movable subarrays.
We begin by presenting a spatially non-stationary channel model to capture the near-field effects, including position-dependent large-scale channel gains and line-of-sight visibility.
To evaluate system performance, we further derive a closed-form approximation of the expected weighted sum rate under maximum ratio combining (MRC), revealing that optimizing XL-MA placement enhances user channel power gain  to increase desired signal power and reduces channel correlation to decreases multiuser interference.
Building upon this, we formulate an antenna placement optimization problem to maximize the expected weighted sum rate, leveraging statistical channel conditions and user distribution. 
To efficiently solve this challenging non-linear binary optimization problem, we propose a polynomial-time successive replacement algorithm.
Simulation results demonstrate that the proposed XL-MA placement strategy achieves near-optimal performance, significantly outperforming benchmark schemes based on conventional fixed-position antennas.

\end{abstract}

\section{Introduction}\label{sec:intro}

The rapid development of wireless communication networks has been driven by advancements in multi-antenna technologies, evolving from multiple-input multiple-output (MIMO) systems in the fourth-generation (4G) networks to massive MIMO in the 5G era, and anticipated to advance into extremely large-scale MIMO (XL-MIMO) systems in the 6G networks \cite{LU2024Tutorial,WANG2024Tutorial}.
While these advancements promise enhanced spatial resolution and spectral efficiency,  the increasing size of antenna arrays and associated radio frequency (RF) chains  significantly increases hardware costs and energy consumption.
To address this issue, antenna selection has been proposed as an alternative solution \cite{sanayei2004antenna}, where  a subset of antenna elements with favorable channel conditions is activated,  at the expense of additional complexity in the required switching circuitry. 
Another promising direction into sparse and modular antenna arrays \cite{li2025sparse,Wang2023SparseArrays} proposes to reduce the number of active antenna elements by evenly distributing them over a given physical area, which creates a large virtual aperture to improve spatial resolution.
However, all the aforementioned technologies rely on fixed-position antennas (FPAs), which cannot fully exploit channel variations in spatial regions and thus limit their communication performance.

Recently, movable antenna (MA) technology  has been regarded as a promising solution to fully exploit channel spatial variations via antenna movement \cite{ZHU2024Movable,ZHU2024Modeling, zhu2025tutorial}, sometimes also referred to as fluid antenna system in terms of flexible antenna positioning \cite{wong2022bruce}.
In contrast to traditional FPA architectures, by dynamically adjusting the positions of MAs within a region, MA-aided wireless systems offer several key advantages, such as signal-to-noise ratio (SNR) boost \cite{ZHU2024Modeling,MEI2024MovableAntennaa}, interference mitigation \cite{zhu2023movable}, flexible beamforming \cite{ma2024multibeam}, and enhanced spatial multiplexing \cite{MA2024MIMO, ZHU2024MovableAntenna,Xiao2024Multiuser}, using the same number of antennas or even fewer compared to conventional FPA systems.
Due to these advantages, extensive studies have delved into the antenna position optimization in various wireless systems. 
For instance, alternating optimization \cite{MA2024MIMO,Tang2025Secure}, successive convex approximation \cite{qin2024antenna,Hu2025Movable}, gradient descent \cite{ZHU2024MovableAntenna,hu2024fluid}, among others, were adopted to search for the locally optimal antenna positions in a continuous space  based on the field-response channel model tailored to MA systems \cite{ZHU2024Modeling}.
To reduce the implementation complexity, a group sparse MA architecture has been proposed in \cite{LU2024Group}, where multiple antennas are concurrently moved with the aid of a single driver. 
To acquire complete channel mappings between transmit and receive regions, compressed sensing-based approaches were proposed by recovering field-response information of channel paths \cite{Ma2023Compressed,Xu2024Channel}.
To reduce overhead from frequent antenna movement and real-time channel estimation, optimizing MA positions based on statistical channel state information (CSI) has been proposed in \cite{ye2023fluid,hu2024two,zheng2024two,zhu2025movable,yan2025movable,shao20246d,SHAO20256Da,SHAO20256DMA}.
For example, statistical-CSI-based MA position optimization has been used to maximize channel capacity in MA-aided MIMO systems \cite{ye2023fluid}, extended to multiuser systems with a two-timescale design under Rician fading channels \cite{hu2024two, zheng2024two}, and also applied to enhance near-field communication performance \cite{zhu2025movable}.
Additionally, \cite{yan2025movable} introduced a statistical field-response channel model incorporating both light-of-sight (LoS) and non-light-of-sight (NLoS) components and  developed a two-timescale optimization framework.
In \cite{shao20246d,SHAO20256Da,SHAO20256DMA}, a six-dimensional (6DMA) system was proposed to optimize 3D positions and 3D rotations of antenna surfaces based on spatial user distribution for maximizing multiuser MIMO capacity.

However, previous studies on MAs have mainly concentrated on wavelength-scale movement regions, where the variations are restricted to phase shifts, providing limited ability to combat large-scale path loss. 
To overcome these limitations, the pinching antenna technique \cite{Ding2025Flexible,liu2025pinching,ouyang2025array} has been recently proposed, which activates antennas by applying small dielectric particles, such as plastic pinches, to a dielectric waveguide.
The flexible positioning of pinching antennas along an extended length allows them to move closer to target users, establishing strong LoS channels and reducing large-scale path loss. 
However, pinching antennas with a shared RF chain can only be moved on a 
preconfigured 1D waveguide, which limits their movement flexibility and spatial multiplexing performance. 
Additionally, existing studies on pinching antenna systems have largely assumed pure LoS channels, which cannot fully capture real-world propagation environments with multipath effects.

To address these issues, in this paper, we propose a new MA architecture, namely extremely large-scale MA (XL-MA), which allows for flexible movement of antennas/subarrays over a 2D extremely large region.
By offering higher degrees of freedom (DoFs) in antenna placement, the XL-MA design significantly enhances the near-field effects of the entire array, thereby reducing multiuser channel correlation and improving spatial multiplexing performance. 
In particular,  we investigate an uplink XL-MA-enabled multiuser communication system, where multiple single-antenna users spatially distributed over a given 3D area are served by a base station (BS) equipped with multiple movable subarrays.
We aim to improve the long-term multiuser communication performance based on statistical channels and/or user distribution via XL-MA placement optimization.
 The contributions of this paper are summarized as follows:
 \begin{itemize}
    \item We develop a 3D grid-based spatial model characterizing  the user distribution, associating each grid with a reference position and a probabilistic indicator. A spatially non-stationary channel model is then presented to capture near-field effects and local propagation non-stationarity, incorporating position-dependent large-scale path gains for both LoS and NLoS components, angles of arrival (AoAs)  with respect to (w.r.t.) antenna and user positions, and LoS visibility. Using these models, the multiuser system performance is evaluated via the expected weighted sum rate of all grids, which involves averaging over random user-activation indicators and channel matrix.

    \item To reduce computational complexity for our design, we derive a closed-form analytical approximation for the expected weighted sum-rate based on the low-complexity maximum ratio combining (MRC) scheme. 
    This approximation significantly reduces computational complexity compared to existing methods for rate approximation such as Monte Carlo simulations and enables efficient optimization of MA subarray placement, balancing performance and complexity. We reveal that XL-MA system can efficiently enhance user channel power gain to increase desired signal power  and reduce channel correlation  to decrease interference, greatly improving multiuser communication performance.

	\item Building upon this approximation, we formulate an optimization problem to maximize the expected weighted sum rate under MRC by determining the optimal MA subarray positions. 
	To efficiently solve the resulting non-linear binary problem, we reformulate it into a reduced-dimensional form and propose a successive replacement algorithm with polynomial complexity by combining high-quality initialization and iterative refinement.

	\item Numerical results show that the proposed algorithm achieves near-optimal performance. Furthermore, the results show that our proposed MRC-based subarray placement optimization approaches the derived performance upper bounds when optimal minimum mean square error (MMSE) combining is employed. Additionally, the proposed XL-MA placement strategy significantly outperforms benchmark schemes based on dense or sparse arrays with FPAs, under both MRC and MMSE combining schemes.

\end{itemize}

The remainder of this paper is organized as follows. Section~\ref{sec:sys} presents the system model, including the XL-MA region, user distribution, and channel modeling. 
Section~\ref{sec:prob}  analyzes the system performance and
formulates the subarray placement optimization problem.
Section~\ref{sec:solution} details the proposed two-stage solution framework. Simulation results are provided in Section~\ref{sec:sim}, followed by conclusions in Section~\ref{sec:Conclusion}.

\textit{Notations}:
$\tbinom{n}{k}$ denotes the number of combinations to choose $k$ elements from a set of $n$ elements.
$\jmath$ denotes the imaginary unit of a complex number.
$x\sim \mathcal{CN}(0, \sigma^2)$ 
represents a circularly symmetric
complex Gaussian random variable $x$ with zero mean and variance $\sigma^2$.
$\mathbb{E}\{\cdot\}$ denotes the statistical expectation.
$(\cdot)^{\sf H}$ and $(\cdot)^{\sf T}$ denote the conjugate transpose and transpose, respectively.
For a complex-valued vector $\bm x$, $\|\bm x\|$ denotes its Euclidean norm. 
$\text{diag}(\bm z)$ denotes a diagonal matrix with each diagonal entry being the corresponding element in vector $\bm z$.
$\bm{x} \circ \bm{z}$ denotes the Hadamard (element-wise) product of two vectors $\bm{x}$ and $\bm{z}$, with elements $[\bm{x} \circ \bm{z}]_n = [\bm{x}]_n \cdot [\bm{z}]_n$, while $\bm{x}^{\circ 2}$ represents the Hadamard square of $\bm{x}$, with elements $[\bm{x}^{\circ 2}]_n = ([\bm{x}]_n)^2$.
For a set $\cal{S}$, $|\cal{S}|$ denotes its cardinality. $\emptyset$ denotes an empty set.
For two sets $\cal{S}$ and $\cal{S}'$, $\cal{S}\cap \cal{S}'$ denotes their intersection, $\cal{S}\cup \cal{S}'$ denotes their union, and $\cal{S}\backslash \cal{S}'$ is the set of elements that belong to $\cal{S}$ but are not in $\cal{S}'$.

\section{System Model}\label{sec:sys}
\begin{figure}[t]
	\centering
	\includegraphics[width=0.9\linewidth]{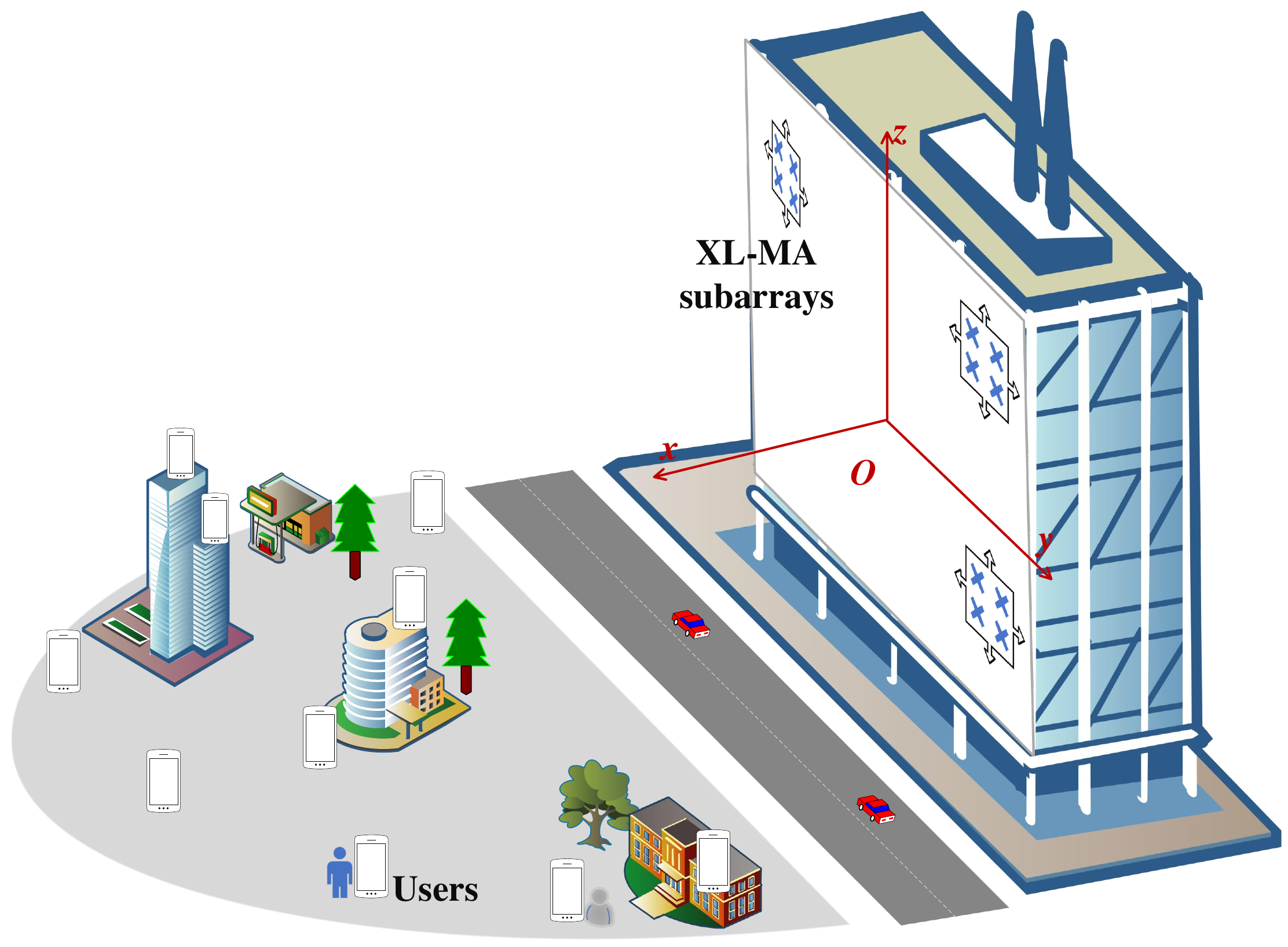} 
	\vspace{-0.5em}	
	\caption{Illustration of an XL-MA-enabled multiuser communication system.}
	\label{fig:system}
	\vspace{-1.5em}
\end{figure}

 As shown in Fig. \ref{fig:system}, we consider an uplink XL-MA-enabled multiuser communication system, where multiple single-antenna users  spatially distributed over a given 3D area (denoted by $\mathcal{C}_{\mathrm{UE}}$) are served by a BS equipped with $N$ MA subarrays. Each MA subarray is a uniform planar array (UPA) of size $M = M_{\mathrm{H}} \times M_{\mathrm{V}}$,
 where $M_{\mathrm{H}}$ and $M_{\mathrm{V}}$ are the number of antennas along the horizontal and vertical directions, respectively, with their respective antenna spacing denoted by $d_{\mathrm{V}}$ and $d_{\mathrm{H}}$.
 With the aid of mechanical drivers, each of the $N$ MA subarrays can be flexibly positioned within a 2D region, denoted by $\mathcal{C}_{\mathrm{MA}}$.
 Different from previous works with wavelength-scale antenna movement (e.g., \cite{MA2024MIMO,ZHU2024MovableAntenna, LU2024Group}), this paper considers an XL-MA region, where the physical size of $\mathcal{C}_{\mathrm{MA}}$ is significantly larger than the  physical aperture of each MA subarray\footnote{The proposed XL-MA system resembles existing BS architectures,  where each MA subarray functions as an active antenna unit (AAU), connected to a distributed unit (DU) and subsequently to a centralized unit (CU) via low-loss optical fibers, enabling long-distance fronthaul transmission and centralized signal processing.}.
 This design with higher DoFs in antenna moving regions can effectively enhance near-field effects of the entire MA array, thereby reducing multiuser channel correlation and  enhancing spatial multiplexing performance. 
 To reduce the overhead of frequently relocating MA subarrays, their positions are optimized in this paper based on statistical channels to improve long-term communication performance, as will be detailed in the following subsections.

\subsection{XL-MA Model}\label{subsec:region}

As shown in Fig. \ref{fig:system}, to describe antenna positions, we establish a 3D coordinate system centered at the BS, where the $y$- and $z$-axes respectively represent the horizontal and vertical directions in the MA region $\mathcal{C}_{\mathrm{MA}}$, and the positive $x$-axis is perpendicular to the antenna moving plane. Without loss of generality, we define the coordinate of each MA subarray's geometric center as its position.

To enable low-cost implementation of antenna movement over such an extremely large area, we discretize the XL-MA region $\mathcal{C}_{\mathrm{MA}}$ by dividing it into a finite number of candidate positions for placing the $N$ MA subarrays.
Specifically, for ease of exposition, we consider that the 2D region $\mathcal{C}_{\mathrm{MA}}$ is a rectangular area in the $y$-$O$-$z$ plane, expressed as
$
\mathcal{C}_{\mathrm{MA}} = [y^{\min}_{\mathrm{MA}}, y^{\max}_{\mathrm{MA}}] \times [z^{\min}_{\mathrm{MA}}, z^{\max}_{\mathrm{MA}}],
$
where $y^{\min}_{\mathrm{MA}}$, $y^{\max}_{\mathrm{MA}}$, $z^{\min}_{\mathrm{MA}}$, and $z^{\max}_{\mathrm{MA}}$ denote the lower and upper bounds of the $y$- and $z$-coordinates, respectively.
To facilitate XL-MA placement, we uniformly sample the region $\mathcal{C}_{\mathrm{MA}}$ into $N_0 = N_y \times N_z$ (with $N_0\gg N$) candidate positions along the $y$- and $z$-axes, where $N_y$ and $N_z$ denote the number of candidate positions along the corresponding axes, respectively.
The 3D coordinate of the $\tilde{n}$-th candidate location is denoted by $\mathbf{r}_{\tilde{n}}\in  \mathbb{R}^{3}$, which is expressed as
\begin{align}
\mathbf{r}_{\tilde{n}} =
\begin{bmatrix}
0 \\
y^{\min}_{\mathrm{MA}} + \left(n_y - \tfrac{1}{2} \right)\Delta_{\mathrm{MA}}^y \\
z^{\min}_{\mathrm{MA}} + \left(n_z - \tfrac{1}{2} \right)\Delta_{\mathrm{MA}}^z
\end{bmatrix}\in  \mathcal{C}_{\mathrm{MA}}, 
1 \leq \tilde{n} \leq  N_0,
\end{align}
where $(n_y, n_z)$ represents the 2D index of the $\tilde{n}$-th candidate position, determined by the linear mapping $\tilde{n} = n_y + (n_z - 1)N_z$, with $1\leq n_y \leq N_y$ and $1\leq n_z \leq N_z$. 
Here,  $\Delta_{\mathrm{MA}}^y \triangleq  \frac{y^{\max}_{\mathrm{MA}}-y^{\min}_{\mathrm{MA}}}{K_y}$ and $ \Delta_{\mathrm{MA}}^z \triangleq  \frac{z^{\max}_{\mathrm{MA}}-z^{\min}_{\mathrm{MA}}}{K_z}$ denote the uniform sampling intervals along the $y$- and $z$-axes, respectively. 
These intervals are chosen to be larger than the physical dimensions of each MA subarray along the corresponding axes, thereby ensuring that any two subarrays placed at adjacent candidate positions do not overlap.

Let $\mathcal{N}_0 = \{1, 2, \ldots, N_0\}$ denote the index set of all candidate positions, and let $\mathcal{N} = \{1, 2, \ldots, N\}$ denote the index set of all MA subarrays. 
We define a selection mapping $\mu: \mathcal{N} \rightarrow \mathcal{N}_0$, where $\mu(n) \in \mathcal{N}_0$ specifies the index of the candidate position selected for placing the $n$-th MA subarray, $\forall n \in \mathcal{N}$. 
Accordingly, the center position of the $n$-th MA subarray is given by $\mathbf{r}_{\mu(n)}, \forall n \in \mathcal{N}$.
To facilitate the candidate position selection for each MA subarray, we define the antenna placement matrix $\bm{\Phi} \in \{0,1\}^{N \times N_0}$, where each element is specified in terms of $\{\mu(n)\}$ as follows:
\begin{align} \label{Eq:Indicator}
[\bm{\Phi}]_{n,\tilde{n}} =
\begin{cases}
1, & \text{if } \tilde{n} = \mu(n), \, \forall n \in \mathcal{N}, \\
0, & \text{otherwise}.
\end{cases}
\end{align}
Accordingly, the matrix $\bm{\Phi}$ is subject to the following constraints to ensure valid subarray placement:
\begin{eqnarray}
    \bm{1}^{\sf T}_{N}\bm{\Phi} &\leq& \bm{1}_{N_0}^{\sf T}, \label{P0Cons2}\\
    \bm{\Phi}\bm{1}_{N_0} &=& \bm{1}_{N}, \label{P0Cons1}
\end{eqnarray}
where constraint \eqref{P0Cons2} ensures that each candidate position accommodates at most one MA subarray, and constraint \eqref{P0Cons1} ensures that each MA subarray is deployed at exactly one candidate position.

\subsection{Grid-Based User Distribution Model}\label{subsec:user}

In this subsection, we develop a 3D grid-based spatial model of user distribution, associating each grid with a reference position and a probabilistic indicator, which preserves large-scale distribution characteristics while enabling analytical tractability for system design and evaluation.

We begin by discretizing the 3D coverage region $\mathcal{C}_{\mathrm{UE}}$ into a finite number of small 3D grids, as in \cite{fu2024multi}.
For analytical tractability, we consider that the 3D region $\mathcal{C}_{\mathrm{UE}}$ is a cuboid\footnote{For irregular geometries of $\mathcal{C}_{\mathrm{UE}}$, we can always find a larger cuboid including it and assume that the user density outside $\mathcal{C}_{\mathrm{UE}}$ is zero.}, expressed as
$
\mathcal{C}_{\mathrm{UE}} = [x^{\min}_{\mathrm{UE}}, x^{\max}_{\mathrm{UE}}] \times [y^{\min}_{\mathrm{UE}}, y^{\max}_{\mathrm{UE}}] \times [z^{\min}_{\mathrm{UE}}, z^{\max}_{\mathrm{UE}}],
$
where $x^{\min}_{\mathrm{UE}}$, $x^{\max}_{\mathrm{UE}}$, $y^{\min}_{\mathrm{UE}}$, $y^{\max}_{\mathrm{UE}}$, $z^{\min}_{\mathrm{UE}}$, and $z^{\max}_{\mathrm{UE}}$ denote its lower and upper bounds of the $x$-, $y$-, and $z$-coordinates, respectively.
To ensure that all users lie in the half-space in front of the MA subarrays, we set $x^{\min}_{\mathrm{UE}} > 0$.
To simplify the representation of the continuous  user locations, we discretize $\mathcal{C}_{\mathrm{UE}}$ into $K = K_x \times K_y \times K_z$ non-overlapping 3D grids along the $x$-, $y$-, and $z$-axes, where $K_x$, $K_y$, and $K_z$ denote the number of grids along the corresponding axes, respectively.
Furthermore, let $\mathbf{t}_{k} \in \mathbb{R}^3$ denote the geometric center of the $k$-th grid, 
which is expressed as
\begin{align}\label{Eq:centerGrid}
\mathbf{t}_{k} =
\begin{bmatrix}
x^{\min}_{\mathrm{UE}} + \left(k_x - \tfrac{1}{2} \right)\Delta_{\mathrm{UE}}^x \\
y^{\min}_{\mathrm{UE}} + \left(k_y - \tfrac{1}{2} \right)\Delta_{\mathrm{UE}}^y \\
z^{\min}_{\mathrm{UE}} + \left(k_z - \tfrac{1}{2} \right)\Delta_{\mathrm{UE}}^z
\end{bmatrix} \in \mathcal{C}_{\mathrm{UE}},  1\leq k\leq K,
\end{align}
where $(k_x, k_y, k_z)$ denotes the 3D index of the $k$-th grid along the $x$-, $y$-, and $z$-axes, respectively, with $1\leq k_x \leq K_x$, $1\leq k_y \leq K_y$, and $1\leq k_z \leq K_z$, and the grid index $k$ is computed as
$k = k_x + (k_y - 1)K_x + (k_z - 1)K_x K_y$.
Here, $ \Delta_{\mathrm{UE}}^x \triangleq  \frac{x^{\max}_{\mathrm{UE}}-x^{\min}_{\mathrm{UE}}}{K_x}$, $ \Delta_{\mathrm{UE}}^y \triangleq  \frac{y^{\max}_{\mathrm{UE}}-y^{\min}_{\mathrm{UE}}}{K_y}$,  and $ \Delta_{\mathrm{UE}}^z \triangleq  \frac{z^{\max}_{\mathrm{UE}}-z^{\min}_{\mathrm{UE}}}{K_z}$ respectively represent the uniform discretization intervals along the $x$-, $y$- and $z$-axes.

We consider in this paper a scenario where multiple users are randomly distributed within the  given 3D service area $\mathcal{C}_{\mathrm{UE}}$.
Due to random channel fading, it is reasonable to assume that each user has an identical probability of being assigned to different time-frequency resource blocks (RBs) \cite{marzetta2016fundamentals}.
Thus, with a sufficiently large number of users and RBs, the spatial user distributions over different RBs are regarded to be statistically identical.  Accordingly, it suffices to focus on the user distribution over a typical RB in the sequel.

Given that the size of each grid is much smaller than the signal propagation distances, the channel vectors for users located within the same grid are expected to exhibit high spatial correlation, which can be attributed to the limited spatial separation among user positions within the grid.
To avoid severe intra-grid interference, it is reasonable to assume that at most one user per grid is assigned to the typical RB \cite{marzetta2016fundamentals}.
To model the user distribution over the RB, we define a random binary variable $\alpha_k\in \{0,1\}$ for grid $k$, $1\leq k \leq K$, where $\alpha_k = 1$ indicates that a user in grid $k$ is assigned to the RB, and $\alpha_k = 0$ otherwise.
Let $\rho_k \in [0,1]$ denote the probability that a user in grid $k$ is assigned to the RB, which is proportional to the user density in that grid.
Then, we have 
\begin{align}
	\mathbb{P}(\alpha_k = 1) = \rho_k, \quad \mathbb{P}(\alpha_k = 0) = 1 - \rho_k, \forall k,
\end{align}
where $\mathbb{P}(\cdot)$ denotes the probability of an event. The expected number of users over this RB, $\bar{K}$, is given by 
\begin{align}\label{Eq:ExpectedK} 
	\bar{K} = \sum_{k = 1}^K \rho_k,
\end{align}
which satisfies $\bar{K} \leq K$.

\subsection{Spatially Non-Stationary Channel Model}\label{subsec:channel}
Next, we characterize the baseband equivalent and spatially non-stationary channel from any user in each grid to the BS.
In a practical multipath propagation environment, the channel from each user and an individual MA subarray comprises  an LoS component (if present) and  NLoS components caused by scattering and reflections.
Since the aperture of each MA subarray is much smaller than the signal propagation distance, any user in each grid $k$ is assumed to be in the far-field region of each individual subarray placed at any candidate position $\tilde{n}$, i.e.,
\begin{align}
\|\bm{r}_{\tilde{n}} - \bm{t}_k\| > \frac{2D^2_{\text{subarray}}}{\lambda}, 1 \leq k \leq K, 1\leq \tilde{n} \leq N,
\end{align}
where $D_{\text{subarray}} \triangleq \sqrt{(M_H -1)^2d^2_H+(M_V -1)^2d^2_V}$ represents the maximum dimension of the subarray aperture with $\lambda$ being the wavelength.
However, due to the extensive aperture created by widely distributed subarrays across the region $\mathcal{C}_{\mathrm{MA}}$, any user in each grid $k$ is located in the near-field region  of the entire XL-MA array, i.e.,  
\begin{align}
	\|\bm{r}_{\tilde{n}} - \bm{t}_k\| < \frac{2D_{\text{XL-MA}}^2}{\lambda}, 1 \leq k \leq K, 1\leq \tilde{n} \leq N,
\end{align}
where $D_{\text{XL-MA}} \triangleq = \max_{ \tilde{n},  \tilde{n}' } \|\bm{r}_{\tilde{n}} - \bm{r}_{\tilde{n}'}\|$ represents the maximum aperture of the region  $\mathcal{C}_{\mathrm{MA}}$. 
The near-field effect introduces spatial variations in key channel parameters across different subarray positions, including the AoAs of the LoS path and the large-scale path gains of both LoS and NLoS components \cite{LU2024Tutorial}. 
Moreover, the spatial non-stationarity is further intensified by variations in local propagation conditions across different positions within $\mathcal{C}_{\mathrm{MA}}$, arising from heterogeneous environmental blockages and the non-uniform distribution of surrounding scatterers.
Consequently, the LoS visibility between any grid and each MA subarray may vary with different antenna positions within $\mathcal{C}_{\mathrm{MA}}$\cite{Carvalho2020NonStationarities}.

Let $\bm{h}_{k} (\{\mathbf{r}_{\mu(n)}\}) \in \mathbb{C}^{NM}$ denote the channel vector from a user in grid $k$ to all the antennas of the $N$ MA subarrays at the BS, which is expressed as
\begin{align} \label{Eq:StackedChannel}
\bm{h}_{k} (\{\mathbf{r}_{\mu(n)}\}) &= 
\begin{bmatrix}
\tilde{\bm{h}}_{k}(\mathbf{r}_{\mu(1)})^{\sf T},
\tilde{\bm{h}}_{k}(\mathbf{r}_{\mu(2)})^{\sf T} ,
\ldots,
\tilde{\bm{h}}_{k}(\mathbf{r}_{\mu(N)})^{\sf T}
\end{bmatrix}^{\sf T}, \nonumber\\
& \hspace{12em}1\leq k \leq K,
\end{align}
where $\tilde{\bm{h}}_{k}(\mathbf{r}_{\mu(n)}) \in \mathbb{C}^{M}$ denotes the channel between any user in grid $k$ and the $n$-th MA subarray placed at $\mathbf{r}_{\mu(n)}$.

Next, we model the channel subvector $\tilde{\bm{h}}_{k}(\mathbf{r}_{\mu(n)})$, $\forall n \in \mathcal{N}$, accounting for the position-dependent LoS path existence and large-scale path loss variations.
First, to represent LoS presence, we introduce a binary LoS visibility indicator $\xi_{\text{LoS}, k}(\mathbf{r}_{\mu(n)})\in \{0,1\}$, where $\xi_{\text{LoS}, k}(\mathbf{r}_{\mu(n)}) = 1$ indicates that an LoS path exists between grid $k$ and the candidate position $\mathbf{r}_{\mu(n)}$, and $\xi_{\text{LoS}, k}(\mathbf{r}_{\mu(n)}) = 0$ otherwise.
Accordingly, the channel $\tilde{\bm{h}}_{k}(\mathbf{r}_{\mu(n)})$ is modeled as
\begin{align} \label{Eq:ChannelModel1}
\tilde{\bm{h}}_{k}(\mathbf{r}_{\mu(n)})  = \xi_{\text{LoS}, k}(\mathbf{r}_{\mu(n)}) \tilde{\bm{h}}_{\text{LoS}, k}(\mathbf{r}_{\mu(n)}) + \tilde{\bm{h}}_{\text{NLoS}, k}(\mathbf{r}_{\mu(n)}), 
\end{align}
where $\tilde{\bm{h}}_{\text{LoS}, k}(\mathbf{r}_{\mu(n)}) \in \mathbb{C}^{M}$ and $\tilde{\bm{h}}_{\text{NLoS}, k}(\mathbf{r}_{\mu(n)}) \in \mathbb{C}^{M}$ denote the LoS and NLoS components, respectively.

\subsubsection{LoS Channel Component}
Under the far-field condition between grid $k$ and the $n$-th MA subarray, the impinging wavefront can be approximated as planar.
Since the position of a user in grid $k$ is random, the grid center $\mathbf{t}_{k}$ is used as the reference position to approximate the large-scale channel characteristics, with the AoA and the LoS path gain remaining approximately constant across the grid.
To specify the direction of arrival of the planar wavefront in Cartesian coordinates, we define the normalized wave vector as
\begin{align}\label{Eq:Channel}
\!\!\!\!	\bm{u}_{k}(\mathbf{r}_{\mu(n)}) = (\mathbf{t}_{k} - \mathbf{r}_{\mu(n)}) / \|\mathbf{t}_{k} - \mathbf{r}_{\mu(n)}\| \in\mathbb{R}^{3\times 1},  \forall k, \forall n.
\end{align}
Then, the UPA receive array response vector of  the $n$-th MA subarray w.r.t. any user in grid $k$ is constructed as the Kronecker product of the vertical and horizontal array response vectors, given by
\begin{align}
	\mathbf{a}(\bm{u}_{k}(\mathbf{r}_{\mu(n)}))   
	 = \mathbf{a}_{M_{V}}(\bm{u}_{k}(\mathbf{r}_{\mu(n)})[3])  \otimes \mathbf{a}_{M_{H}}(\bm{u}_{k}(\mathbf{r}_{\mu(n)})[2]), \nonumber
\end{align}
where $\mathbf{a}_{M'}(u') = [e^{-\jmath\frac{2\pi}{\lambda}\Delta'(m-1)u'} ]_{m=1}^{M'}\in \mathbb{C}^{M'}$, with $\lambda$ and $\Delta'$ denoting the carrier wavelength and inter-element spacing, respectively.
Let $\beta_{\mathrm{LoS},k}(\mathbf{r}_{\mu(n)}) \in \mathbb{R}$ denote the LoS path gain (if any) from grid $k$ to position $\mathbf{r}_{\mu(n)}$, which depends on their distance.
Accordingly, the LoS component is modeled as   
\begin{align}\label{Eq:LoS}
	\tilde{\bm{h}}_{\text{LoS},k}(\mathbf{r}_{\mu(n)}) &= \sqrt{\beta_{\mathrm{LoS},k}(\mathbf{r}_{\mu(n)})} e^{-\jmath \psi_{k, \mu(n)}} \mathbf{a}(\bm{u}_{k}(\mathbf{r}_{\mu(n)})), \nonumber\\
 &\hspace{7em} 1\leq k \leq K, 1\leq n \leq N,
\end{align}
where  $\psi_{k, \mu(n)} \in \mathbb{C}$  denotes the random phase shift  caused by the randomness of user location within grid $k$, which is modeled as a random variable uniformly distributed within $[0, 2\pi)$, i.e., $\psi_{k, \mu(n)} \sim \mathcal{U} [0, 2\pi)$.

\subsubsection{NLoS Channel Components}
In the presence of rich scatterers in the environment (e.g., buildings and vehicles), the NLoS components between a user in grid $k$ and the $n$-th MA subarray are modeled as uncorrelated Rayleigh fading.
Let $\beta_{\mathrm{NLoS},k}(\mathbf{r}_{\mu(n)}) \in \mathbb{R}$ denote the average NLoS path gain from any position in grid $k$ to position $\mathbf{r}_{\mu(n)}$, which depends on the distance $\|\mathbf{t}_{k} - \mathbf{r}_{\mu(n)}\|$.
The NLoS channel is thus modeled as
\begin{align}\label{Eq:NLoS}
	\tilde{\bm{h}}_{\text{NLoS}, k}(\mathbf{r}_{\mu(n)}) &\sim \mathcal{CN}(0, \beta_{\text{NLoS}, k}(\mathbf{r}_{\mu(n)}) \mathbf{I}_{M}),  \forall k, \forall n.
\end{align}

Building upon the above, we denote the virtual channel matrix from all the $K$ grids to all candidate positions of the XL-MA array at the BS by
\begin{align}
\bar{\bm{H}} = [  \bar{\bm{h}}_{1}, \bar{\bm{h}}_{2},\ldots,  \bar{\bm{h}}_{K}     ]	\in \mathbb{C}^{N_0M\times K},
\end{align}
where $\bar{\bm{h}}_{k} = [ (\tilde{\bm{h}}_{k}(\mathbf{r}_1))^{\sf T},  (\tilde{\bm{h}}_{k}(\mathbf{r}_2))^{\sf T},\ldots,  (\tilde{\bm{h}}_{k}(\mathbf{r}_{N_0}))^{\sf T}  ]^{\sf T} \in \mathbb{C}^{N_0M}$, $1 \leq k\leq K$, is the channel vector from the user in grid $k$ to all candidate positions  of the MA subarray. 
It can be shown that $\bm{h}_{k} (\{\mathbf{r}_{\mu(n)}\})$ in \eqref{Eq:StackedChannel} can also be represented in terms of $\bm{\bar H}$ as 
\begin{align}
	\bm{h}_{k} (\{\mathbf{r}_{\mu(n)}\}) =  (\bm {\Phi} \otimes \mathbf{I}_{M})\bm{\bar H} \bm{e}_k =  (\bm {\Phi} \otimes \mathbf{I}_{M})\bm{\bar h}_k, \forall k,
\end{align}
where  $\bm{e}_k \in \mathbb{R}^{K}$ denotes the $k$-th canonical basis vector (i.e., the vector whose $k$-th entry is $1$ and all other entries are $0$).

\subsection{Uplink Transmission}\label{subsec:metric}

Based on the above models, the received uplink signal at the BS over the typical RB is given by
\begin{align}
	\bm y    & =  (\bm {\Phi} \otimes \mathbf{I}_{M})\bm{\bar H} \bm{A} \bm{P}^{1/2} \bm {s} + \bm{\omega},
\end{align}
where $\bm{A} \triangleq \diag(\alpha_1, \alpha_2,\ldots,   \alpha_K )\in \mathbb{R}^{K\times K}$ is a diagonal matrix formed by the random binary indicator vector $\bm{\alpha} = [\alpha_1, \alpha_2, \ldots, \alpha_K]^{\sf T}\in  \mathbb{R}^{K}$, $\bm {s} \triangleq [s_1, s_2,\ldots,  s_K ]^{\sf T}\in \mathbb{C}^{K}$, with $s_k \in \mathbb{C}$ denoting the symbol transmitted by a user in grid $k$ (if any) with $\mathbb{E}\{|s_k|^2\} = 1$, and $\bm {P}^{1/2} \triangleq \diag(\sqrt{P_1}, \sqrt{P_2},\ldots,  \sqrt{P_K} )\in \mathbb{R}^{K\times K}$ is a diagonal matrix with $P_k\in\mathbb{R}$ denoting the corresponding transmit power.
In addition, $\bm{\omega} \in \mathbb{C}^{NM \times 1}$ denotes the additive white Gaussian noise (AWGN) vector at the BS, which is modeled as $\bm{\omega} \sim \mathcal{CN}(\mathbf{0}, \sigma^2 \mathbf{I}_{NM})$.

For a given XL-MA position selection matrix $\bm {\Phi}$,  and conditioned on $\alpha_k = 1$, i.e., a user in grid $k$ is assigned over the typical RB, let $\bm{v}_k\in \mathbb{C}^{NM}$ denote the instantaneous combining vector applied at the BS for a user in grid $k$, $1\leq k \leq K$, which is dependent on the channel matrix $(\bm {\Phi} \otimes \mathbf{I}_M)\bm{\bar H}_0 \bm{A}$.
After applying the combining vector $\bm{v}_k$, the combined signal from the user in grid $k$ is given by
\begin{align}
	\bar{y}_k = \bm{v}_k^{\sf H}\bm{y} 
	&= \underbrace{\bm{v}_k^{\sf H}(\bm {\Phi} \otimes \mathbf{I}_{M})\bar{\bm{h}}_{k} \sqrt{P_k} s_k}_{\text{Desired signal}} \nonumber \\
	&+\underbrace{\sum_{{i=1, i\neq k}}^K  \bm{v}_k^{\sf H}(\bm {\Phi} \otimes \mathbf{I}_{M})\bar{\bm{h}}_{i} \alpha_i\sqrt{P_i} s_i}_{\text{Interference}} + \bm{v}_k^{\sf H}\bm{\omega}.
\end{align}
Accordingly, the instantaneous signal-to-interference-plus-noise ratio (SINR) for a user in grid $k$ is given by
\begin{align}\label{Eq:SINR1}
	\gamma_k = \frac{\bar{P}_k \left|\bm{v}_k^{\sf H} (\bm {\Phi} \otimes \mathbf{I}_{M})\bar{\bm{h}}_{k} \right|^2}{\sum_{{i=1, i\neq k}}^K \bar{P}_i \alpha_i\left| \bm{v}_k^{\sf H} (\bm {\Phi} \otimes \mathbf{I}_{M})\bar{\bm{h}}_{i} \right|^2 + \|\bm{v}_k\|^2}, 
\end{align} 
where $\bar{P}_k \triangleq \frac{P_k}{\sigma^2}$ represents the transmit SNR. 
As such, conditioned on $\alpha_k = 1$, the expected achievable rate of a user in grid $k$ is given by
\begin{align}\label{Eq:Rate1}
	R_k(\bm {\Phi}) &= \mathbb{E}_{\bm{\alpha}_{-k},\bm H}\left\{\log_2(1+  \gamma_k)\right\},
\end{align}
where the expectation $\mathbb{E}_{\bm{\alpha}_{-k}, \bm H}\left\{\cdot\right\}$ is taken w.r.t. the random binary sub-vector $\bm{\alpha}_{-k} \in \{0,1\}^{K-1}$, obtained by removing the $k$-th element from $\bm{\alpha}$, and the random channel matrix $\bm{H} \triangleq (\bm{\Phi} \otimes \mathbf{I}_M)\bar{\bm{H}}$.
Note that since $\bm{v}_k$ changes with $(\bm{\Phi} \otimes \mathbf{I}_M)\bm{\bar H}_0 \bm{A}$,  the expectation in \eqref{Eq:Rate1} implicitly  averages over  $\bm{v}_k$.

Building upon the above, the overall system performance over this typical RB is evaluated by the expected weighted sum rate, given by 
\begin{align}\label{Eq:AverageSumRate}
	R_{\text{sum}}(\bm {\Phi}) 
	= \sum_{k = 1}^{K} \mathbb{P}(\alpha_k = 1) R_k(\bm {\Phi})
	= \sum_{k = 1}^{K} \rho_k R_k(\bm {\Phi}).  
\end{align}

The expectation operator $\mathbb{E}_{\bm{\alpha}_{-k}, \bm{H}}\{\cdot\}$ in $R_k(\bm{\Phi})$ involves averaging over both the random binary indicators $\bm{\alpha}_{-k}$ and the  channel matrix $\bm{H}$, which is challenging to directly compute in closed form.
One possible approach is to use Monte Carlo simulations to approximate the expectation, but this can be computationally expensive.
To tackle these challenges, the next section derives a closed-form approximation for $R_k(\bm{\Phi})$ in \eqref{Eq:Rate1} based on the low-complexity MRC scheme.
Moreover, MRC is well-suited for XL-MA systems with a limited number of antennas, as the large antenna moving region allows optimized subarray positions to reduce user channel correlation, achieving a balance between performance and complexity.
It is important to note that MRC is only adopted to assist MA position optimization. 
In practice, once the MA subarrays are deployed to the optimized positions, more advanced combining schemes, such as MMSE combining \cite{bjornson2017massive}, can be employed based on instantaneous channels, similar to conventional FPA systems.

\section{Performance Analysis and Problem Formulation}\label{sec:prob}

 In this section, we first present the closed-form expression of the expected weighted sum rate under the MRC receiver. Then, we formulate the  optimization problem of MA subarray placement  for expected weighted sum-rate maximization.

\subsection{Expected Rate Approximation Under MRC}\label{subsec:MRCapprox}

By substituting MRC vectors, i.e., $\bm{v}_k = (\bm {\Phi} \otimes \mathbf{I}_{M})\bar{\bm{h}}_{k}, 1\leq k \leq K,$ into \eqref{Eq:Rate1}, the expected achievable rate for a user in grid $k$
is simplified as
\begin{align}\label{Eq:RateMR}
	R_k^{\text{MR}}(\bm {\Phi}) = 
	 \mathbb{E}_{\bm{\alpha}_{-k}, \bm H}\left\{\log_2\left(1+ \gamma_k^{\text{MR}}\right)\right\},  \forall k,
\end{align}
with
\begin{align} 
	 	\gamma_k^{\text{MR}} = \frac{\bar{P}_k \left|\bm{\bar h}^{\sf H}_{k} (\bm {\Phi}^{\sf H}\bm {\Phi} \otimes \mathbf{I}_{M})\bm{\bar h}_{k}\right|^2}{\sum\limits_{\substack{i=1,\\i\neq k}}^K \bar{P}_i \alpha_i\left|\bm{\bar h}^{\sf H}_{k} (\bm {\Phi}^{\sf H}\bm {\Phi} \otimes \mathbf{I}_{M})\bm{\bar h}_{i}\right|^2 + \bm{\bar h}^{\sf H}_{k} (\bm {\Phi}^{\sf H}\bm {\Phi} \otimes \mathbf{I}_{M})\bm{\bar h}_{k}}, \nonumber
\end{align}
where $(\bm {\Phi} \otimes \mathbf{I}_{M})^{\sf H} (\bm {\Phi} \otimes \mathbf{I}_{M}) = (\bm {\Phi}^{\sf H}\bm {\Phi} \otimes \mathbf{I}_{M})$.

In the following theorem, we provide a closed-form approximation for the expected achievable rates under MRC  receiver.

\begin{theorem}\label{theorem:MRCapprox}
Using MRC receiver, the expected achievable rate for a user in grid $k$ defined in \eqref{Eq:RateMR} can be approximated by
\begin{align} \label{Neq:RateMRapprox}
	&\tilde{R}^{\text{MR}}_{k} (\bm {\Phi}) 
	=  \log_2\left( 1 +  \tilde{\gamma}_k^{\text{MR}}\right), \forall k,
\end{align}
with 
\begin{align} \label{Neq:sinrMRapprox}
	&\tilde{\gamma}_k^{\text{MR}} = \nonumber\\
	&\frac{\bar{P}_k\Big (M^2 \left( \bm{1}_N^{\sf T}\bm{\Phi} \bm{\bar \beta}_{k} \right)^2 + M \bm{1}_N^{\sf T}\bm{\Phi}(\bm{\bar{\beta}}_k^{\circ 2} \circ\bm{\bar{f}}_k) \Big)}
	{ \bm{1}_N^{\sf T}\bm{\Phi}\Big( \sum_{\substack{i=1, \\ i\neq k}}^K \bar{P}_i \rho_i \bm{\bar \beta}_{k}\circ\bm{\bar{\beta}}_i \circ (\bm{\bar{\varphi}}_{k,i}\circ \bm{\bar{g}}_{k,i} + \bm{\bar{q}}_{k,i}) + M\bm{\bar \beta}_{k}\Big)},
\end{align}
where  $\bm{\bar{\beta}}_k \in \mathbb{R}^{N_0}$ and $\bm{\bar{\varphi}}_{k,i}\in \mathbb{R}^{N_0}$   are respectively defined as
\begin{align} \label{Eq:totalgain}
	[\bm{\bar{\beta}}_k]_{\tilde{n}} = \xi_{\text{LoS},k}(\mathbf{r}_{\tilde{n}}) \beta_{\text{LoS},k}(\mathbf{r}_{\tilde{n}}) + \beta_{\text{NLoS},k}(\mathbf{r}_{\tilde{n}}),   \forall\tilde{n} \leq N_0,
\end{align}
and 
\begin{align}
	[\bm {\bar \varphi}_{k,i}]_{\tilde{n}} &= \frac{\sin^2(M_{V}\pi\frac{d_{V}}{\lambda}(\bm{u}_{k}(\mathbf{r}_{\tilde{n}})[3]-\bm{u}_{i}(\mathbf{r}_{\tilde{n}})[3]))}{\sin^2(\pi\frac{d_{V}}{\lambda}(\bm{u}_{k}(\mathbf{r}_{\tilde{n}})[3]-\bm{u}_{i}(\mathbf{r}_{\tilde{n}})[3]))} \times\nonumber\\
	& \frac{\sin^2(M_{H}\frac{\pi d_{H}}{\lambda}(\bm{u}_{k}(\mathbf{r}_{\tilde{n}})[2]-\bm{u}_{i}(\mathbf{r}_{\tilde{n}})[2]))}{\sin^2(\frac{\pi d_{H}}{\lambda}(\bm{u}_{k}(\mathbf{r}_{\tilde{n}})[2]-\bm{u}_{i}(\mathbf{r}_{\tilde{n}})[2]))},  \forall \tilde{n}.
\end{align}
In addition, the auxiliary vectors $\bm{\bar{f}}_k \in \mathbb{R}^{N_0}$, $\bm{\bar{g}}_{k,i} \in \mathbb{R}^{N_0}$, and $\bm{\bar{q}}_{k,i} \in \mathbb{R}^{N_0}$ are respectively defined in \eqref{Eq:fdefine}, \eqref{Eq:gdefine}, and \eqref{Eq:qdefine} in Appendix \ref{theorem:MRCapprox_proof}.
\end{theorem}

\noindent\textit{Proof}: See Appendix \ref{theorem:MRCapprox_proof}.

Therefore, using MRC, the expected weighted sum rate,  $R^{\text{MR}}_{\text{sum}}(\bm {\Phi}) \triangleq \sum_{k = 1}^{K} \rho_k R_k^{\text{MR}}(\bm {\Phi})$,
 is approximated as
\begin{align}\label{Eq:AverageSumRateApprx}
	R^{\text{MR}}_{\text{sum}}(\bm {\Phi}) 
	\approx 
	\sum_{k = 1}^{K} \rho_k \tilde{R}^{\text{MR}}_{k} (\bm {\Phi}).  
\end{align}

As observed from \eqref{Neq:RateMRapprox} and \eqref{Neq:sinrMRapprox},  
the expected achievable rate of the user in grid $k$ is primarily governed by the average signal power and interference power, which are characterized by large-scale channel conditions.
On the one hand, the average signal power of the user in grid $k$ increases with its large-scale path gains  $\bm{\bar{\beta}}_k$, each determined by the LoS indicator $\xi_{\text{LoS},k}(\mathbf{r}_{\tilde{n}})$ and the LoS/NLoS path gains $\beta_{\text{LoS},k}(\mathbf{r}_{\tilde{n}})$ and $\beta_{\text{NLoS},k}(\mathbf{r}_{\tilde{n}})$. To maximize the average signal power, it is desirable to deploy each MA subarray at a candidate position $\mathbf{r}_{\tilde{n}}$  where $\xi_{\text{LoS},k}(\mathbf{r}_{\tilde{n}}) = 1$ and path gains are maximized.

On the other hand, the average interference power depends on the large-scale path gains of the desired user ($\bm{\bar{\beta}}_k$) and interfering users ($\bm{\bar{\beta}}_i$, $\forall i \neq k$), the correlation between their channels ($\bm{\bar{\varphi}}_{k,i}$), and the activation probabilities of interfering users ($\rho_i$, $\forall i \neq k$).
To suppress interference, one strategy is to place the MA subarrays at positions $\mathbf{r}_{\tilde{n}}$ where the large-scale path gains of interfering users are minimized, i.e., $\xi_{\text{LoS},i}(\mathbf{r}_{\tilde{n}}) = 0$ and/or the path gains are minimized, leveraging spatially non-stationary LoS characteristics.
However, this may weaken the link quality of the interfering users in other grids, degrading their achievable rates $\tilde{R}^{\text{MR}}_{i}(\bm{\Phi})$'s and the overall weighted sum rate $\sum_{k = 1}^{K} \rho_k \tilde{R}^{\text{MR}}_{k}(\bm{\Phi})$, especially when $\rho_i$ is large.
This highlights a trade-off between interference suppression and signal enhancement: aggressively mitigating interference by avoiding strong interfering links may compromise the desired signal power for other users, ultimately degrading overall system performance.
Beyond the signal power domain, interference mitigation can also be achieved by reducing the channel correlation between the desired and interfering users via XL-MA placement optimization.
Since the antenna moving region is sufficiently large, the spatial resolution of the entire XL-MA array can be effectively improved due to the near-field effects such that users with small angle/distance difference can be distinguished. 
In summary, XL-MA systems can significantly increase user channel power and decrease users' channel correlation via XL-MA placement optimization, thereby improving multiuser communication performance significantly.

\subsection{Problem Formulation}\label{subsec:prob}

Building upon the above analysis, we formulate the optimization problem for determining the optimal positions of $N$ MA subarrays, which aims to maximize the expected weighted sum rate under practical constraints.
The associated optimization problem is formulated as 
\begin{subequations}  \label{Problem1}
\begin{align}
	 \text{(P1):~} \mathop{\text{maximize}}\limits_{ \bm{\Phi}} \quad &   \sum_{k = 1}^{K} \rho_k  \tilde{R}^{\text{MR}}_{k} (\bm{\Phi})  \label{P0obj}\\
	\text{subject to} \quad
	&[\bm{\Phi}]_{n,\tilde{n}} \in \{0,1\}, \forall n \in \mathcal{N},  \forall \tilde{n} \in \mathcal{N}_0, \\
	& \eqref{P0Cons2} \text{~and~}\eqref{P0Cons1}. \label{P1:Cons2} 
\end{align}	
\end{subequations}

(P1) is a binary non-linear optimization problem with a non-concave objective function, which is challenging to be optimally solved using standard convex optimization techniques.
Moreover, since $\tilde{R}^{\text{MR}}_{k} (\bm{\Phi})$ in \eqref{P0obj} depends only on the selection indictor vector $\bm{1}_N^{\sf T}\bm{\Phi}$, the order of the $N$ selected positions does not affect the objective value, making the optimal solution to (P1) non-unique.
One straightforward approach to optimally solve (P1) is by using the exhaustive search method, which evaluates all possible combinations of $N$ candidate positions out of the $N_0$ candidate positions. 
However, this method is computationally prohibitive, especially when $N_0$ and/or $N$ are large, as it requires evaluating $\binom{N_0}{N}$ combinations. 
To overcome this challenge, we propose in Section \ref{sec:solution} a suboptimal algorithm to solve (P1) efficiently.

\section{Proposed Solution to (P1)}\label{sec:solution}

In this section, we first reformulate (P1) into a more tractable form to facilitate efficient optimization. Subsequently, we propose a successive replacement algorithm with a polynomial computational complexity to solve the reformulated problem effectively.

\subsection{Problem Reformulation}\label{subsec:reformulation}
To facilitate efficient algorithm and reduce the number of optimization variables, we introduce a new optimization variable vector $\bm{\chi} \in \mathbb{R}^{N_0}$, defined as  $ \bm{\chi}^{\sf T} \triangleq \bm{1}_N^{\sf T}\bm{\Phi}$.
For notational simplicity, we henceforth write $\tilde{R}^{\text{MR}}_{k} (\bm{\Phi})$ as $\tilde{R}^{\text{MR}}_{k} (\bm{\chi})$, by replacing $\bm{1}_N^{\sf T}\bm{\Phi}$ with $\bm{\chi}$ in the expression of $\tilde{R}^{\text{MR}}_{k} (\bm{\Phi})$ given in \eqref{Neq:RateMRapprox}.
By introducing the variable transformation, constraint \eqref{P0Cons2} in (P1) is inherently satisfied due to the binary nature of $\bm{\chi}$, while constraint \eqref{P0Cons1} in (P1) is equivalently reformulated by the linear constraint $\bm{1}_{N_0}^{\sf T}\bm{\chi} = N$.
As a result, (P1) can be equivalently reformulated in terms of the variable $\bm{\chi}$ as follows:
\begin{subequations}  
\begin{align}
		 \text{(P2):~} \mathop{\text{maximize}}\limits_{ \bm{\chi}} \quad &   \sum_{k = 1}^{K} \rho_k  \tilde{R}^{\text{MR}}_{k} ( \bm{\chi})  \label{P2:obj}\\
		\text{subject to} \quad
		& \bm{1}_{N_0}^{\sf T}\bm{\chi} = N, \label{P2:Cons4} \\
		&[ \bm{\chi}]_{\tilde{n}} \in \{0,1\},   \forall \tilde{n} \in \mathcal{N}_0,
\end{align}	
\end{subequations}
where constraint \eqref{P2:Cons4} ensures that exactly $N$ candidate positions are selected for placing MA subarrays.
It is evident that the optimal objective values of (P1) and (P2) are identical. 
However, problem (P2) remains a binary non-linear optimization problem with a non-concave objective, making it intractable to solve optimally due to the logarithmic structure and the coupling of multiple selected MA positions, which cannot be linearized in terms of $\bm{\chi}$ in general.
To efficiently solve (P2), the next subsection introduces a linear surrogate function for effective position initialization, followed by a successive replacement stage to iteratively refine MA subarray positions and improve the objective value.

\subsection{Proposed Successive Replacement Algorithm}

 \subsubsection{Initialization}\label{subsec:initialization}

 To obtain a high-quality initial positions of MA subarrays, we simplify (P2) into a linear optimization problem, ensuring that grids with high user densities/probabilities have LoS paths (if any) to at least one MA subarray.   
 Specifically,   we adopt a linear surrogate by approximating each term $\tilde{R}^{\text{MR}}_{k}(\bm{\chi})$ in \eqref{P2:obj} as the sum of marginal contributions from individual MA subarrays.  The marginal contribution of a subarray at position $\mathbf{r}_{\tilde{n}}$ to grid $k$ is quantified as the expected rate when only that subarray is active, corresponding to the selection vector $\bm{\chi} = \bm{e}_{\tilde{n}}$, where $\bm{e}_{\tilde{n}}$ is the canonical basis vector.
 Accordingly, by substituting $\bm{\chi} = \bm{e}_{\tilde{n}}$ into $\tilde{R}^{\text{MR}}_{k}(\bm{\chi})$ in \eqref{Neq:RateMRapprox}, the resulting rate reduces to
\begin{align} \label{Neq:rateMRapprox1}
	& \tilde{R}^{\text{MR}}_{k}(\bm{e}_{\tilde{n}})=\nonumber\\
	& \log_2\Big(1+\frac{\bar{P}_kM [\bm{\bar \beta}_{k}]_{\tilde n}\big (M   +  [\bm{\bar{f}}_k]_{\tilde n} \big)}
	{ \sum_{\substack{i=1, \\ i\neq k}}^K \bar{P}_i \rho_i  [\bm{\bar{\beta}}_i ]_{\tilde n} ([\bm{\bar{\varphi}}_{k,i}]_{\tilde n} [\bm{\bar{g}}_{k,i}]_{\tilde n} + [\bm{\bar{q}}_{k,i}]_{\tilde n}) + M}\Big), \nonumber\\
	& \hspace{18em} \forall k, \forall \tilde{n},
\end{align}
As such, the sum of marginal contributions to grid $k$, denoted by $\hat{R}_{k}(\bm{\chi})$, are computed as
\begin{align} \label{Eq:rateMRapprox3}
\hat{R}_{k}(\bm{\chi}) = \sum_{\tilde{n} = 1}^{N_0}[\bm{\chi}]_{\tilde{n}} \tilde{R}^{\text{MR}}_{k}(\bm{e}_{\tilde{n}}), \forall k.
\end{align}
Note that $\hat{R}_{k}(\bm{\chi})$ in \eqref{Eq:rateMRapprox3} is a linear function of $\bm{\chi}$, which still preserves key channel characteristics, such as position-specific path gains and user channel correlations, crucial for identifying optimal MA subarray positions.

In addition, it is observed from \eqref{P2:obj} that the expected weighted sum-rate, increases linearly with $\rho_k,\forall k$.
Improving the data rate of high user-density grids thus significantly improves the overall system performance. As noted in Section~\ref{subsec:MRCapprox}, maximizing $\tilde{R}^{\text{MR}}_{k} (\bm{\chi})$ requires deploying each MA subarray at a candidate position $\mathbf{r}_{\tilde{n}}$ where $\xi_{\text{LoS},k}(\mathbf{r}_{\tilde{n}}) = 1$.
Motivated by this, the initialization ensures that each of the top-$N$ grids with the highest user densities/probabilities is covered by at least one MA subarray via an LoS link, if such a link exists, i.e., $\sum_{{\tilde n} \in \mathcal{N}_0} \xi_{\text{LoS},k}(\mathbf{r}_{\tilde{n}}) \geq 1$. Let $\mathcal{K}_N = \{k_1, k_2, \ldots, k_N\}$ denote the indices of the top $N$ grids, ordered such that
$\rho_{k_1} \geq \rho_{k_2} \geq \cdots \geq \rho_{k_N} > 0$,
where $k_1, k_2, \ldots, k_N$ are distinct and satisfy $\rho_k \geq \rho_i$ for all $k \in \mathcal{K}_N$ and $i \in \{1,2,\ldots, K\} \setminus \mathcal{K}_N$.  
If fewer than $N$ grids have non-zero user-density/probability values, we set $\mathcal{K}_N = \{k \in \{1,\ldots,K\}\mid \rho_k >0\}$.

Building on the above, by further relaxing each entry of $\bm{\chi}$ to be continuous within $[0,1]$, the associated optimization problem can be formulated as 
\begin{subequations} \label{Problem4}
\begin{align}
	\text{(P2.1):~}
\mathop{\text{maximize}}_{\bm{\chi}} \quad
&  \sum_{k=1}^{K} \rho_k \hat{R}_{k}(\bm{\chi}) \label{P3:obj} \\
\text{subject to}  \quad
&\sum_{{\tilde n}=1}^{N_0} [\bm{\tilde{\chi}}]_{\tilde n} \xi_{\text{LoS},k}(\mathbf{r}_{\tilde{n}})  \geq 1,\forall k \in   \mathcal{\hat K}, \label{P3:Cons1}\\
& \bm{1}_{N_0}^{\sf T}\bm{\chi} = N, \label{P3:Cons2}\\
& 0 \le [\bm{\tilde{\chi}}]_{\tilde n}\le 1, \forall {\tilde n} \in \mathcal{N}_0, \label{P3:Cons3}
\end{align}
\end{subequations}
where $\mathcal{ \hat K} \triangleq \{k\in \mathcal{K}_N \mid  \sum_{{\tilde n} \in \mathcal{N}_0} \xi_{\text{LoS},k}(\mathbf{r}_{\tilde{n}}) \geq  1\}$ denotes the set of top-$N$ grids that are LoS-reachable by at least one candidate position.
(P2.1) is a convex linear programming (LP), which can be solved optimally by standard solvers. 

Let $\bm{\chi}^*$ denote its optimal solution.
The index of the candidate position selected for placing the $n$-th MA subarray, denoted by $\mu(n)$, corresponds to the index of the $n$-th largest entry in $\bm{\chi}^*$.
Accordingly, the index set of the selected $N$ candidate positions is given by
\begin{align}\label{Eq:initial}
\mathcal{N}_{\mu} = \{\mu(1),\mu(2),\ldots, \mu(N)\},
\end{align}
such that $[\bm{\chi}^*]_{\mu(1)}\geq [\bm{\chi}^*]_{\mu(2)} \geq \cdots \geq [\bm{\chi}^*]_{\mu(N)}$, with $\mu(n)\neq \mu(n')$ for $n\neq n'$, and  $[\bm{\chi}^*]_{\tilde{n}} \geq [\bm{\chi}^*]_{\tilde{n}'}$ for all $\tilde{n}\in \mathcal{N}_{\mu}$, $\tilde{n}' \in \mathcal{N}_0 \setminus \mathcal{N}_{\mu}$.

\subsubsection{Successive Replacement}\label{subsec:successive}

We can construct a feasible solution $\bm{\Phi}$ to (P1) by substituting $\mathcal{N}_{\mu}$ into \eqref{Eq:Indicator}, but it may not be optimal due to the linear surrogate in \eqref{Eq:rateMRapprox3}.
 To further enhance performance, we propose a successive replacement procedure that iteratively replaces one selected candidate position in $\mathcal{N}_{\mu}$ with an unselected one from $\mathcal{N}_0 \setminus \mathcal{N}_{\mu}$, while keeping the remaining selected positions fixed.
 Specifically, let $\mathcal{N}_{\text{repl}} \subseteq \mathcal{N}$ denote the set of MA subarrays whose positions have already been replaced in the first $(t-1)$ iterations. At iteration $t$, we aim to improve performance by updating the position $\mu(n_t) \in \mathcal{N}_{\mu}$, for deploying the $n_t$-th MA subarray, where $n_t \in \mathcal{N} \setminus \mathcal{N}_{\text{repl}}$. The rule for selecting the index $n_t$ of the MA subarray to update at iteration $n$ is specified later.
 Given this setup, the optimization problem at the $t$-th iteration is formulated as
 \begin{subequations}
	\begin{align}
	(\text{P}2.2.t:)~
	\mathop{\text{maximize}}_{\tilde{n}_t,\bm{\chi}} \quad
	&  \sum_{k=1}^{K} \rho_k \tilde{R}^{\text{MR}}_{k} ( \bm{\chi}) \label{P2.2t:obj} \\
	\text{subject to}  \quad
	&[\bm{\chi}]_{\tilde n} = 1, \forall {\tilde n}\in  \mathcal{N}_{\mu}^{t}\cup \{\tilde{n}_t\},\\
	&[\bm{\chi}]_{\tilde{n}'} = 0, \forall {\tilde n}'\in  \mathcal{N}_{0} \setminus ( \mathcal{N}_{\mu}^{t}\cup \{\tilde{n}_t\}),\nonumber\\
	& \tilde{n}_t \in \mathcal{N}_0 \setminus\mathcal{N}_{\mu}, 
	\end{align}
	\end{subequations}
 where $ \mathcal{N}_{\mu}^{t} = \mathcal{N}_{\mu}\setminus\{\mu(n_t)\}$ denotes the set of selected candidate positions except the position $\mu(n_t)$.
The selection indicator vector $\bm{\chi}$ in (P2.2.$t$) is determined by the index set $\mathcal{N}_{\mu}^{t}\cup {\tilde{n}_t}$, where $\mathcal{N}_{\mu}^{t}$ is fixed. Thus, solving (P2.2.$t$) reduces to finding $\tilde{n}_t \in \mathcal{N}_0 \setminus \mathcal{N}_{\mu}^{t}$ that maximizes the expected weighted sum-rate in \eqref{P2.2t:obj}.
This is achieved by enumerating all $(N_0-N)$ unselected  positions in $\mathcal{N}_0 \setminus \mathcal{N}_{\mu}^{t}$.
Let $\tilde{n}_t^*$ denote the optimal index of (P2.2.$t$), and $\bm{\chi}_t^*$ be the corresponding selection indicator vector.
We then evaluate whether the replacement leads to performance improvement by checking if $\sum_{k=1}^{K} \rho_k \tilde{R}^{\text{MR}}_{k} (\bm{\chi}_t^{*}) >  \sum_{k=1}^{K} \rho_k \tilde{R}^{\text{MR}}_{k} (\bm{\chi}^{*})$ holds.
If this condition is satisfied, we update the $n_t$-th element in $\mathcal{N}_{\mu}$ by setting
\begin{align}
\mu(n_t) \leftarrow \tilde{n}_t^*,
\end{align}
and accordingly update
\begin{align}
\bm{\chi}^* \leftarrow \bm{\chi}_t^*, \quad 
\mathcal{N}_{\text{repl}} \leftarrow \mathcal{N}_{\text{repl}} \cup \{n_t\}.
\end{align}
Otherwise, the successive replacement process terminates. 
It is evident that the objective value of (P2), i.e., the expected weighted sum-rate, increases with each iteration of the replacement process.
Furthermore, the position of each MA subarray is allowed to be updated at most once, leading to an upper bound of $N$ iterations, i.e., $t\leq N$.

Nonetheless, the overall performance of the successive replacement method depends critically on the update order of MA subarray positions in $\mathcal{N} \setminus \mathcal{N}_{\text{repl}}$. 
To address this, we propose a greedy selection strategy that selects the next subarray to update by identifying the one whose temporary removal causes the least performance loss.
Specifically, at the $t$-th iteration, we solve the following problem:
\begin{subequations}
   \begin{align}
 (\text{P}2.3.t):~
   \mathop{\text{maximize}}_{n,\bm{\chi}} \quad
   &  \sum_{k=1}^{K} \rho_k \tilde{R}^{\text{MR}}_{k} ( \bm{\chi}) \label{P2.3t:obj} \\
   \text{subject to}  \quad
   &[\bm{\chi}]_{\tilde n} = [\bm{\chi}^*]_{\tilde n}, \forall {\tilde n}\in  \mathcal{N}_{0}\setminus\{\mu(n)\},\nonumber\\
   &[\bm{\chi}]_{\mu(n)} = 0, \\
   & n \in \mathcal{N} \setminus \mathcal{N}_{\text{repl}}. 
   \end{align}
   \end{subequations}
   This allows us to evaluate the impact of removing each MA subarray placed at the current selected position $\mu(n)$ and rank them accordingly. 
   Similar to (P2.2.$t$), we enumerate all $(N - t + 1)$ MA-subarray indices in $\mathcal{N} \setminus \mathcal{N}_{\text{repl}}$ and compute the corresponding objective value of (P2.3.$t$), denoted by $R_n$.
   The index of the MA subarray to be updated at the $t$-th iteration is then selected as
  $ n_t = \arg\max_{n \in \mathcal{N} \setminus \mathcal{N}_{\text{repl}}} R_n$.

\begin{algorithm}[!t]
	\caption{Successive Replacement Algorithm for (P1)}
	\label{alg:SequentialUpdate}
	\begin{algorithmic}[1]
	\STATE \textbf{Input}:  $N$, $\{\rho_k\}$.
	\STATE \textbf{Initialization}: Obtain $\bm{\chi}^*$ and $\mathcal{N}_{\mu}$ in \eqref{Eq:initial} via solving (P2.1).
	\STATE 
		Initialize $\mathcal{N}_{\text{repl}} = \emptyset$ and set $t = 1$.
		\WHILE {$t \leq N$}
			\STATE Obtain $n_t$ by solving (P2.3.$t$).
			\STATE Obtain $\tilde{n}_t^*$ and $\bm{\chi}_t^{*}$ by solving (P2.2.$t$).
			\IF {$\sum_{k=1}^{K} \rho_k \tilde{R}^{\text{MR}}_{k} (\bm{\chi}_t^{*}) >  \sum_{k=1}^{K} \rho_k \tilde{R}^{\text{MR}}_{k} (\bm{\chi}^{*})$}
				\STATE $\mu(n_t) \leftarrow \tilde{n}_t^*$, $\bm{\chi}^* \leftarrow \bm{\chi}_t^*$, 
				and $\mathcal{N}_{\text{repl}} \leftarrow \mathcal{N}_{\text{repl}} \cup \{n_t\}$.
				\STATE $t \leftarrow t + 1$.
			\ELSE
				\STATE \textbf{break}
			\ENDIF
		\ENDWHILE
		\STATE Construct  $\bm{\Phi}^*$ using $\mathcal{N}_{\mu}$ according to \eqref{Eq:Indicator}.
	\STATE \textbf{Output}: $\bm{\Phi}^*$, $\bm{\chi}^*$, and $\mathcal{N}_{\mu}$.
	\end{algorithmic}
\end{algorithm}

\subsection{Computational Complexity Analysis}
The main procedures of the proposed successive replacement algorithm are summarized in Algorithm \ref{alg:SequentialUpdate}.
Problem (P2.1) is a convex LP that can be solved in polynomial time, with a worst-case complexity of $\mathcal{O}(N^{3.5})$ using the interior-point method.
Then, the algorithm performs up to $N$ iterations in the successive replacement procedure, where solving (P2.2.$t$) takes $\mathcal{O}((N_0 - N) \cdot C)$ and solving (P2.3.$t$) takes $\mathcal{O}(N \cdot C)$ per iteration.
Here, $C$ denotes the cost of evaluating the objective value, which  can be approximated as $\mathcal{O}(K)$. 
Therefore, each iteration requires $\mathcal{O}(N_0 C)$ time, and the overall complexity of the replacement procedure is $\mathcal{O}(NN_0K)$.
Finally, the total complexity of Algorithm~\ref{alg:SequentialUpdate}  is on the order of
$\mathcal{O}(N^{3.5} + NN_0K)$, with a polynomial scaling w.r.t. the system parameters.
As will be shown in Section~\ref{sec:sim}, it achieves near-optimal performance with significantly reduced computational cost as compared to the exhaustive search method.

\begin{figure*}[t]
    \centering
	\subfigure[Expected weighted sum rate vs. $M_{\mathrm{H}}$.]{
        \includegraphics[width=0.31\linewidth]{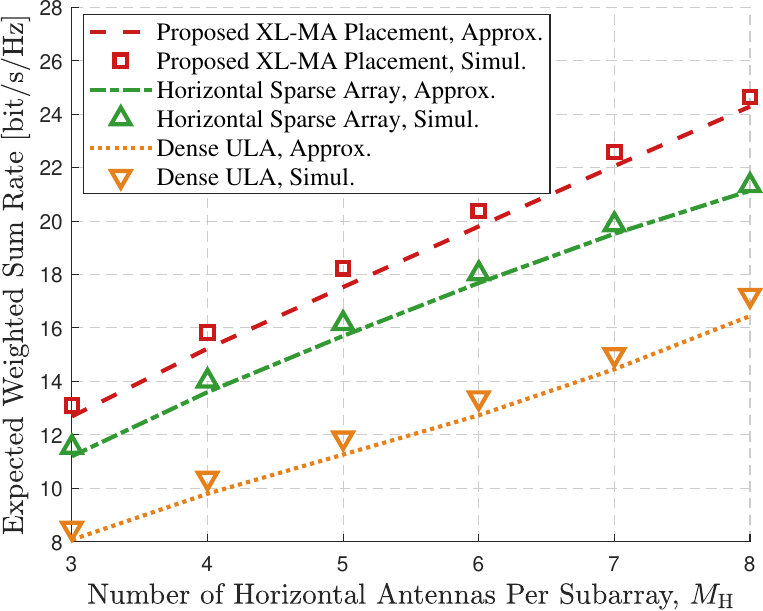} 
        \label{fig:Scenario1tightness}}
	\hfill
	\subfigure[Expected weighted sum rate vs. $y_{\mathrm{MA}}$.]{
        \includegraphics[width=0.31\linewidth]{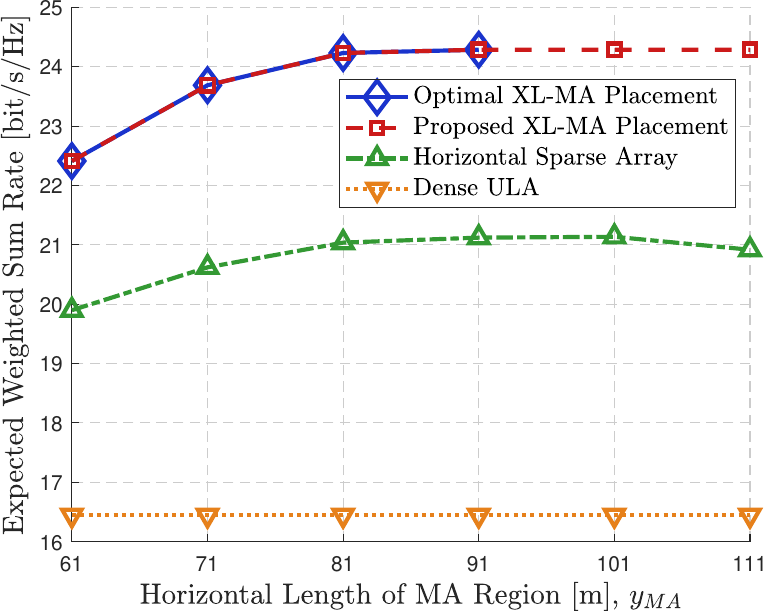} 
        \label{fig:Scenario1movesize}}
	\hfill
	\subfigure[Expected weighted sum rate vs. $\bar{K}$.]{
        \includegraphics[width=0.31\linewidth]{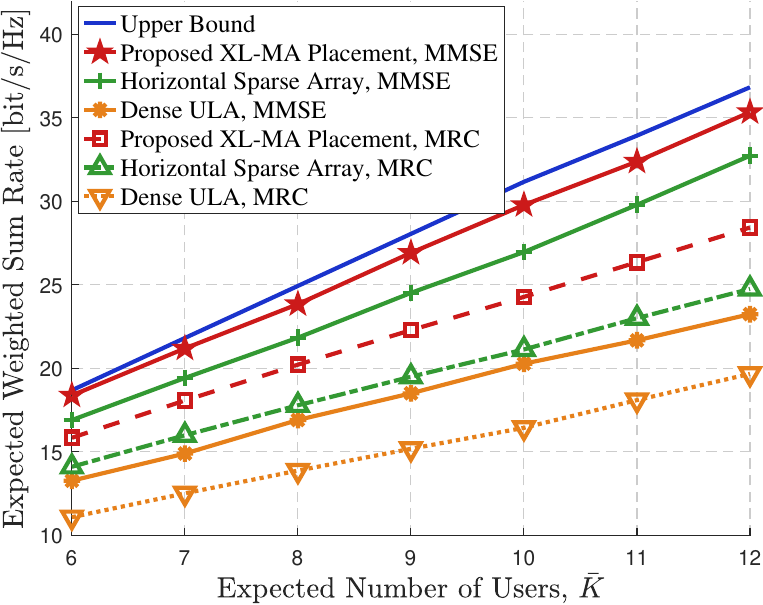} 
        \label{fig:Scenario1NumUE}}
		 \vspace{-0.5em}
    \caption{Performance comparison under full LoS visibility with $\zeta = 0$.}
    \label{fig:Scenario1Performanc}
    \vspace{-1.5em}
\end{figure*}

 \section{Simulation Results}\label{sec:sim}

Simulation results are presented in this section to evaluate the performance of our proposed XL-MA placement solutions by the suboptimal successive replacement algorithm (i.e., Alg. \ref{alg:SequentialUpdate}) under different scenarios, offering valuable insights into their practical applications.

Unless otherwise specified, we consider a scenario with a 2D MA region $\mathcal{C}_{\mathrm{MA}}$ and a 3D coverage region $\mathcal{C}_{\mathrm{UE}}$.
First, the 2D MA region $\mathcal{C}_{\mathrm{MA}}$ is set with $-y^{\min}_{\mathrm{MA}} = y^{\max}_{\mathrm{MA}} = y_{\mathrm{MA}}/2 = 50.5$ m, $z^{\min}_{\mathrm{MA}} = 20$ m, and $z^{\max}_{\mathrm{MA}} = 50$ m, discretized into $N_0 = 3030$ candidate positions with $N_y = 101$ and $N_z = 30$.
Furthermore, the 3D coverage region $\mathcal{C}_{\mathrm{UE}}$ is set with $y^{\max}_{\mathrm{UE}} = - y^{\min}_{\mathrm{UE}} = 52.5$ m, and $x^{\min}_{\mathrm{UE}} = 7.5$ m, $x^{\max}_{\mathrm{UE}} = 52.5$ m, and $z^{\min}_{\mathrm{UE}} = 0$ m, and  $z^{\max}_{\mathrm{UE}} = 50$ m, discretized into $K = 1890$ grids with $K_x = 9$, $K_y = 21$, and $K_z = 10$.
Each grid is assigned a user activation probability: $\hat{\rho}_0$ for regular grids and $\hat{\rho}_1$, $\hat{\rho}_2$ for hotspot regions with higher user density, corresponding to the sets $\mathcal{K}_0$, $\mathcal{K}_1$, and $\mathcal{K}_2$, respectively.
Given the expected number of users $\bar{K}$ and the regular grid ratio $\zeta \in [0,1]$, the three types of probabilities grids are set to satisfy the following constraints:
{\small
\begin{align} \label{Eq:Prob}
	\hat{\rho}_w =
	\begin{cases}
	 \min\{1, \frac{3 \bar{K}(1-\zeta)}{2|\mathcal{K}_1|+3|\mathcal{K}_2|}\}, & w = 2,\\
	 \max\{\frac{2 \bar{K}(1-\zeta)}{2|\mathcal{K}_1|+3|\mathcal{K}_2|}, \frac{\bar{K}(1-\zeta)-|\mathcal{K}_2|}{|\mathcal{K}_1|}\}, & w = 1,\\
	 \frac{\bar{K}\zeta}{|\mathcal{K}_0|}, &  w = 0,
	\end{cases}
	\end{align}
}where $\hat{\rho}_1 \leq \hat{\rho}_2$ holds true.

In the simulations, the number of MA subarrays is set to $N = 8$,
and the antenna spacing is set to $d_V = d_H = \frac{\lambda}{2}$.
The expected number of users is set to $\bar{K} = 10$, the regular grid ratio is set to $\zeta = 0$, and the length of the MA region is set to $y_{\mathrm{MA}} = 101$ m.
If an LoS path exists between grid $k$ and candidate position $\tilde{n}$, the large-scale LoS path gain is given by $\beta_{\mathrm{LoS},k} (\mathbf{r}_{\tilde{n}}) = (\frac{\lambda}{4\pi\|\mathbf{t}_{k} - \mathbf{r}_{\tilde{n}}\|})^2,\forall k, \forall \tilde{n}$.
The average NLoS path gain is set as $\beta_{\mathrm{NLoS},k}(\mathbf{r}_{\tilde{n}}) = \beta_{\mathrm{LoS},k}(\mathbf{r}_{\tilde{n}})/\kappa$, where $\kappa$ is the Rician factor.
Unless otherwise specified, the Rician factor is set to $\kappa \to \infty$, due  to the significant propagation/penetration loss of high-frequency signals.
In addition,  the carrier frequency is set as $f_c = 30$ GHz, and the wavelength is given by $\lambda = \frac{c}{f_c}$, where $c$ is the speed of light.
The transmit power of each user is set to $P_k = 5$ dBm, and the noise power is set to $\sigma^2 = -80$ dBm.
All Monte Carlo simulation results are obtained by averaging over $1000$ independent realizations of random binary grid indictors (i.e, $\bm{\alpha}$) and random channel matrix (i.e., $\bm{H}$).

We evaluate the performance of the proposed XL-MA placement by Alg. \ref{alg:SequentialUpdate} against the following benchmark schemes:
\begin{itemize}

    \item \textit{Sparse Array}: The 8 subarrays are arranged in a $2 \times 4$ configuration along the $z$-axis and $y$-axis, respectively, with positioning index fixed by $\mu(n) =  \lfloor \frac{N_y - 1}{3} \cdot ((n - 1)\bmod 5) + 1 \rceil + \lfloor \frac{n}{5} \rfloor (N_z-1)N_z,\forall n \in \mathcal{N}$.

   \item \textit{Horizontal Sparse Array}: Subarrays are placed along the $y$-axis, with positioning index fixed by $\mu(n) =  \lfloor \frac{N_y - 1}{N-1} \cdot (n - 1) + 1 \rceil,\forall n \in \mathcal{N}$.

   \item \textit{Vertical Sparse Array}: 
	Subarrays are placed along the $z$-axis, with positioning index fixed by $\mu(n) =  (\lfloor \frac{N_z - 1}{N-1} \cdot (n - 1) + 1 \rceil-1)N_z + \lfloor \frac{N_y+1}{2}\rfloor,\forall n \in \mathcal{N}$.

    \item \textit{Dense ULA}: A conventional dense ULA is deployed, consisting of $MN$ antennas arranged along the $y$-axis with antenna spacing of $d_{\mathrm{H}} = \lambda/2$. The array's center is positioned at the candidate location indexed by $\lceil N_0/2 \rceil$.

	\item \textit{Dense UPA}: A conventional dense UPA is employed, where $2M_{\mathrm{V}}$ and $4M_{\mathrm{H}}$ antennas are arranged along the $z$-axis and the $y$-axis, respectively, with antenna spacing of $\lambda/2$. The positioning index is the same as the dense ULA scheme.
\end{itemize}

For ease of demonstration, in Sections \ref{sec:fullLOS} and \ref{subsec:PartialLoS}, we first consider the simple scenario of 1D XL-MA placement and 2D user distribution. 
Specifically, the vertical coordinates of the MA region and user distribution are fixed at  $z^{\min}_{\mathrm{MA}} = z^{\max}_{\mathrm{MA}} = 20.5$ m and $z^{\min}_{\mathrm{UE}} = z^{\max}_{\mathrm{UE}} = 0$ m. 
We set $\mathcal{K}_1 = \{93,99,154,163,172,185\}$, $\mathcal{K}_2 =\{1,9,10,25,28,40\}$, and  $\mathcal{K}_0 = \mathcal{K}\setminus (\mathcal{K}_1\cup \mathcal{K}_2)$.
In this scenario, each subarray is configured as a ULA with $M_{\mathrm{H}} = 8$ and $M_{\mathrm{V}} = 1$. For performance comparison, we evaluate the proposed scheme against two benchmark placements: the horizontal sparse array and the dense ULA.

\begin{figure*}[t]
    \centering
	\subfigure[Proposed XL-MA placement.]{
        \includegraphics[width=0.32\linewidth, height=0.3\linewidth]{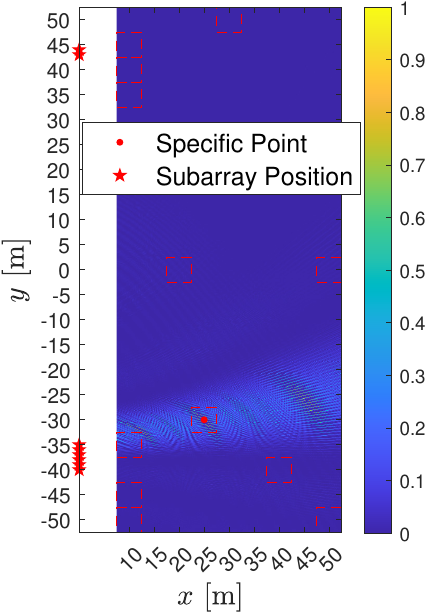} 
        \label{fig:Scenario1proposed}
    }
	\hspace{-1em}
    \subfigure[Horizontal sparse array.]{
        \includegraphics[width=0.32\linewidth, height=0.3\linewidth]{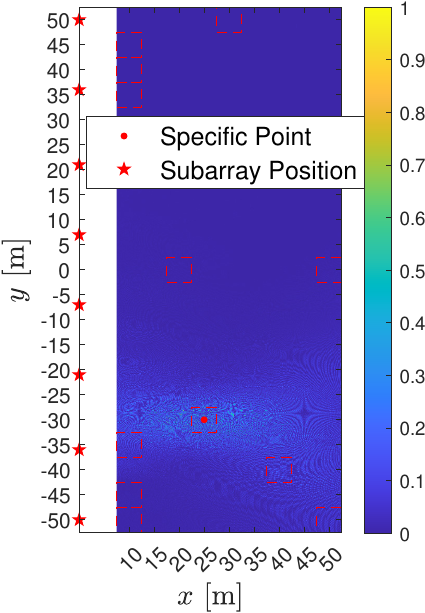} 
        \label{fig:Scenario1Sparse}
    }
	\hspace{-1em}
    \subfigure[Dense ULA.]{
        \includegraphics[width=0.32\linewidth, height=0.3\linewidth]{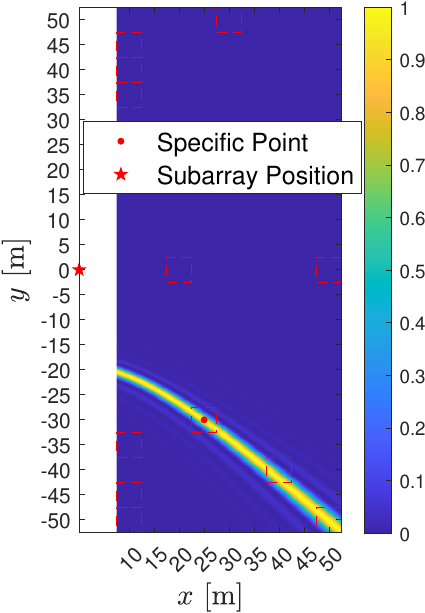} 
        \label{fig:Scenario1Dense}
    }
	\vspace{-1em}
    \caption{Normalized channel correlation of different schemes under full LoS visibility.}
    \label{fig:Scenario1beam}
    \vspace{-1em}
\end{figure*}

\captionsetup{justification=centering}
\begin{figure*}[t]
    \centering
	\subfigure[Proposed XL-MA placement.]{
        \includegraphics[width=0.31\linewidth, height=0.3\linewidth]{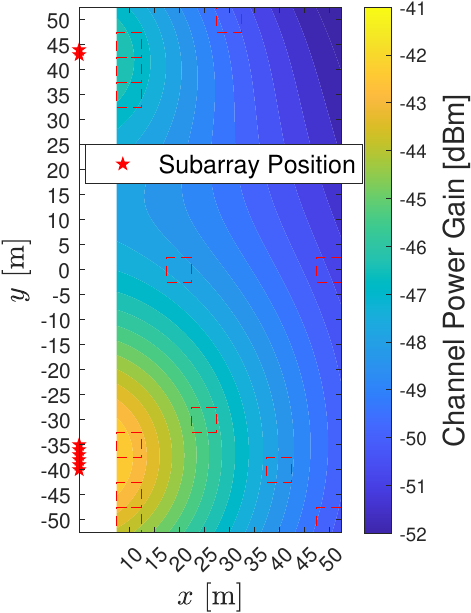} 
        \label{fig:Scenario1proposedChannelGain}
    }
	\hspace{-0.5em}
    \subfigure[Horizontal sparse array.]{
        \includegraphics[width=0.31\linewidth, height=0.3\linewidth]{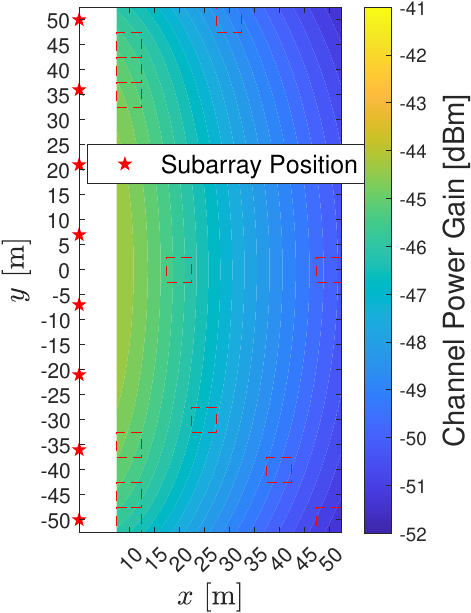} 
        \label{fig:Scenario1SparseChannelGain}
    }
	\hspace{-0.5em}
    \subfigure[Dense ULA.]{
        \includegraphics[width=0.31\linewidth, height=0.3\linewidth]{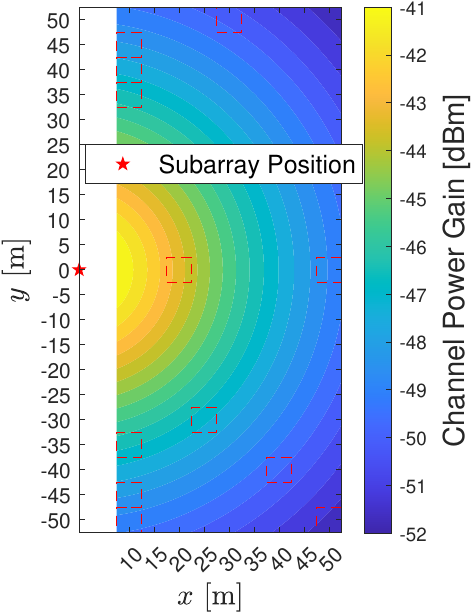} 
        \label{fig:Scenario1DenseChannelGain}
    }
	\vspace{-1em}
    \caption{Channel power gain maps of different schemes under full LoS visibility.}
    \label{fig:Scenario1ChannelGain}
    \vspace{-1.5em}
\end{figure*}

\subsection{Performance Evaluation Under  Full LoS Visibility} \label{sec:fullLOS}

In this subsection, we evaluate the performance of the proposed XL-MA placement scheme in comparison with benchmark schemes under the full LoS visibility scenario, where all candidate positions maintain LoS connectivity with all grids (i.e., $\xi_{\text{LoS},k}(\mathbf{r}_{\tilde{n}}) = 1, \forall k, \forall \tilde{n}$).

Fig.~\ref{fig:Scenario1tightness} shows the expected weighted sum rate under MRC for various MA and FPA schemes as $M_{\mathrm{H}}$ varies. 
For each scheme, the ``Approx.'' curve uses the closed-form analytical approximation (i.e., $\sum_{k = 1}^{K} \rho_k \tilde{R}^{\text{MR}}_{k} (\bm {\Phi})$), while the ``Simul.'' curve averages $R^{\text{MR}}_{\text{sum}}(\bm {\Phi})$ via Monte Carlo simulations. 
It can be observed that the performance by each scheme improves with the increasing $M_{\mathrm{H}}$, due to the enhanced beamforming gain. 
Notably, the proposed XL-MA scheme significantly outperforms the two FPA benchmarks, thanks to its flexible subarray placement that effectively increases the channel power gain of users and decreases their channel correlation.              
Additionally, the close match between the simulation and the analytical results across all schemes confirms the tightness of the approximation in \eqref{Eq:AverageSumRateApprx}. 
Accordingly, all subsequent performance evaluations involving MRC are conducted using this approximation.

Fig.~\ref{fig:Scenario1movesize} shows the expected weighted sum-rate under MRC versus the length of 1D moving segment, i.e., $y_{\mathrm{MA}}$.
The ``Optimal XL-MA Placement" represents the performance by using exhaustive search to solve (P1), with results shown up to $y_{\mathrm{MA}} = 91$ m due to the computational infeasibility of larger regions. 
First, it is observed that the proposed XL-MA placement achieves near-optimal performance, effectively balancing performance and complexity.
Second, the results validate again that the proposed XL-MA placement  outperforms both FPA schemes.
Additionally, as $y_{\mathrm{MA}}$ increases, the proposed scheme's performance improves due to enhanced near-field effects before saturating, while the horizontal sparse array shows non-monotonic performance, with initially increasing rates (from 61 m to 101 m), followed by deceasing rates  due to overly sparse spacing and increased path loss in hotspot regions.

Fig. \ref{fig:Scenario1NumUE} shows the expected weighted sum-rate under MRC and MMSE for various MA and FPA schemes as $\bar{K}$ varies.
In particular, for each placement scheme,  the ``MMSE'' curve is computed by using MMSE combining vectors \cite{bjornson2017massive}  via Monte Carlo simulation technique. 
If interference between users is zero, the expected rate in \eqref{Eq:Rate1} is upper-bounded by $\log_2( 1 + \bar{P}_kM\bm{1}^{\sf T}_N\bm{\Phi}\bm{\bar{\beta}}_k), \forall k$. The ``Upper Bound'' curve is obtained by computing the weighted sum based on this upper bound across all grids for the proposed XL-MA placement.
First, it is observed that the expected weighted sum rate under both MRC and MMSE for all schemes increases with $\bar{K}$, owing to the enhanced spatial multiplexing gain from serving more users concurrently. 
Second, the proposed XL-MA placement scheme outperforms both benchmark schemes under MRC and MMSE receivers, with a growing performance gap as $\bar{K}$ increases, showing its superiority in multiuser communications.
Additionally, the proposed XL-MA placement, optimized based on MRC, achieves an MMSE weighted sum rate close to the theoretical upper bound, demonstrating its efficiency in boosting user channel power gain and reducing channel correlation.


Fig. \ref{fig:Scenario1beam} and Fig. \ref{fig:Scenario1ChannelGain} present the normalized channel correlation and channel power gain maps, respectively.
In each figure, the dashed boxes mark grids with non-zero user densities/probabilities, while the red stars indicate the subarray positions in the 2D top-down view.
First, as shown in Figs.~\ref{fig:Scenario1beam}(a) and \ref{fig:Scenario1ChannelGain}(a), different from the benchmarks, the proposed XL-MA placement groups the subarrays into two clusters: six near grids with the highest user density and two near grids with lower user density, exhibiting a non-uniform sparse geometry.
Next, in Fig. \ref{fig:Scenario1beam}, the normalized channel correlation is calculated as the squared magnitude of the inner product between the normalized channel vector at other points within the coverage region ($ (\bm {\Phi} \otimes \mathbf{I}_{M})\bm{\bar h}_k/\|(\bm {\Phi} \otimes \mathbf{I}_{M})\bm{\bar h}_k\|$) and that at a specified point, which is set to be  $x = 22.5$ m and $y = - 30$ m, marked by the red dot.
As observed in Fig.~\ref{fig:Scenario1beam}(a), the proposed XL-MA placement achieves low channel correlation with the adjacent lower-left grid and negligible correlation with other grids with non-zero user densities/probabilities.
In contrast, the horizontal sparse array in Fig.~\ref{fig:Scenario1beam}(b) and the dense ULA in Fig.~\ref{fig:Scenario1beam}(c) exhibit high correlation between the specified point and its surrounding grids, with the dense ULA showing the strongest correlation with the adjacent lower-right grids.

Moreover, in Fig.~\ref{fig:Scenario1ChannelGain}, we show the channel power gain maps under different schemes, which are obtained by computing the channel gain $\|\left(\bm{\Phi} \otimes \mathbf{I}_M\right)\bar{\bm{h}}_k\|^2$ over all positions within the coverage region $\mathcal{C}_{\mathrm{UE}}$.  
As shown in Fig.~\ref{fig:Scenario1ChannelGain}(a),  the proposed XL-MA placement forms a non-uniform sparse geometry, predominantly concentrating channel power gain on the major hotspot grids.
By contrast, the horizontal sparse array in Fig.~\ref{fig:Scenario1ChannelGain}(b) exhibits a relatively uniform channel power distribution across the coverage area, lacking the focused power enhancement in hotspot regions.
The dense ULA in Fig.~\ref{fig:Scenario1ChannelGain}(c) exhibits strong gain near the center; however, it experiences significant attenuation in key hotspot grids, such as the lower-bottom grids.
These observations show that the proposed XL-MA placement reduces user channel correlation (Fig.~\ref{fig:Scenario1beam}(a)) while ensuring balanced signal power coverage in major hotspot grids, thus enhancing multiuser communication performance.

\begin{figure*}[t]
    \centering
	\subfigure[Proposed XL-MA placement.]{
        \includegraphics[width=0.31\linewidth, height=0.3\linewidth]{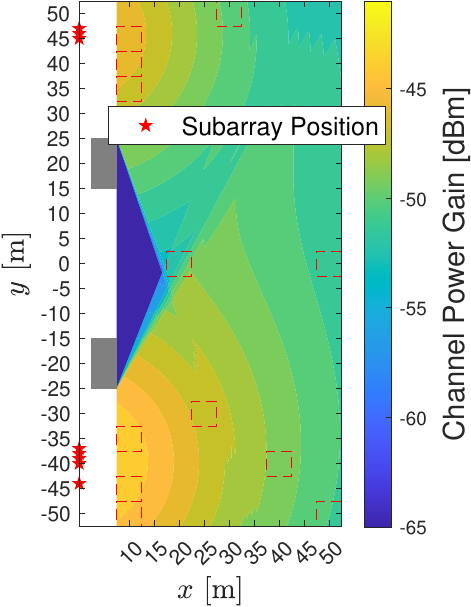} 
        \label{fig:Scenario2proposedChannelGain}
    }
	\hspace{-0.5em}
    \subfigure[Horizontal sparse array.]{
        \includegraphics[width=0.31\linewidth, height=0.3\linewidth]{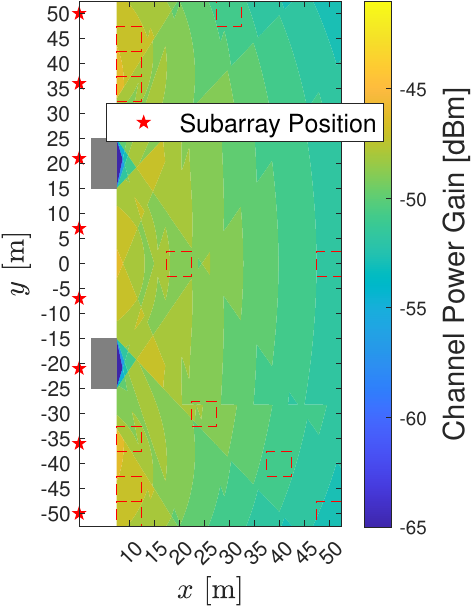} 
        \label{fig:Scenario2SparseChannelGain}
    }
	\hspace{-0.5em}
    \subfigure[Dense ULA.]{
        \includegraphics[width=0.31\linewidth, height=0.3\linewidth]{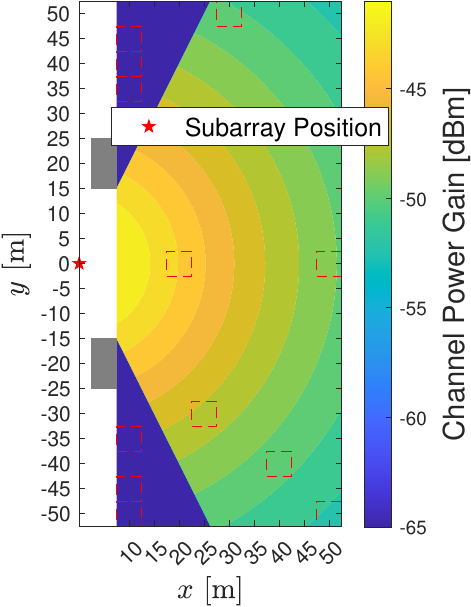} 
        \label{fig:Scenario2DenseChannelGain}
    }
	\vspace{-0.5em}
    \caption{Channel power gain maps of different schemes under partial LoS visibility with $\zeta = 0$ and $\kappa \to \infty$.}
    \label{fig:Scenario2ChannelGain}
    \vspace{-1.5em}
\end{figure*}

\begin{figure}[t]
    \centering
        \includegraphics[width=0.8\linewidth]{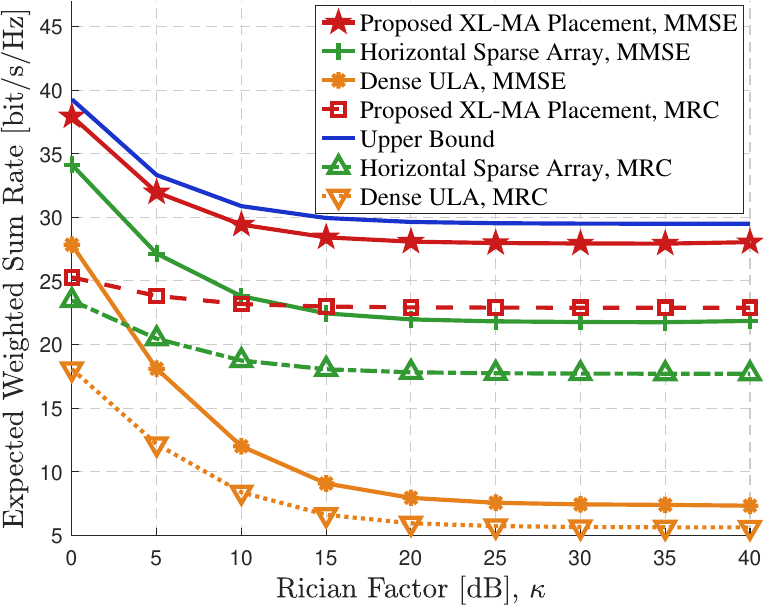} 
		\vspace{-0.5em}
    \caption{Expected weighted sum rate vs. $\kappa$.}
    \label{fig:Scenario2Rician}
    \vspace{-1.5em}
\end{figure}

\subsection{Performance Evaluation Under Partial LoS Visibility}\label{subsec:PartialLoS}
In this subsection, we evaluate the performance of the proposed XL-MA placement scheme against benchmark schemes under a partial LoS visibility scenario, where two obstacles are placed between the movable region and the coverage region. Specifically, the first obstacle is centered at $(5, -20, 9)$ with dimensions $5\,\text{m} \times 10\,\text{m} \times 18\,\text{m}$, and the second is centered at $(5, 20, 9)$ with the same dimensions.
An LoS link between candidate subarray position $\tilde{n}$ and grid $k$ exists (i.e., $\xi_{\text{LoS}, k}(\mathbf{r}_{\tilde{n}}) = 1, \forall k, \forall \tilde{n}$) if the line segment connecting $\mathbf{r}_{\tilde{n}}$ to each of 20 randomly sampled points within the grid does not intersect any obstacle.

Fig.~\ref{fig:Scenario2ChannelGain} presents the channel power gain maps for different schemes under the partial LoS visibility scenario with $\bar{K} = 10$ and $\kappa \to \infty$. The two obstacles are depicted as gray rectangular areas in the 2D top-down view. For consistency across schemes, the LoS path gain is set to $\beta_{\mathrm{LoS},k}(\tilde{n}) = -65$ dBm for any grid $k$ where the LoS path from candidate position $\tilde{n}$ is blocked.
As shown in Fig.~\ref{fig:Scenario2ChannelGain}(a), the optimized subarray positions of the proposed XL-MA placement are again grouped into two geometrically distinct clusters, each located near one of the two user-dense regions, similar to the full LoS visibility case in Fig.~\ref{fig:Scenario1ChannelGain}(a). However, their positions and allocation are strategically adjusted to account for the obstacle-induced LoS blockage. Specifically, five subarrays in the lower cluster are sparsely distributed around the grids associated with $\hat{\rho}_2$, covering both $\hat{\rho}_2$ grids and the middle two $\hat{\rho}_1$ grids. The remaining three subarrays are placed in the upper cluster to cover the $\hat{\rho}_1$ grids. This layout ensures strong LoS coverage for all hotspot regions.
In contrast, Fig.~\ref{fig:Scenario2ChannelGain}(b) shows that the horizontal sparse array, despite broader coverage, suffers from power degradation in critical user regions due to limited LoS links with at most three subarrays per hotspot grid.
The dense ULA in Fig.~\ref{fig:Scenario2ChannelGain}(c) performs worse, with its compact layout causing severe power loss in the lower-left and upper-left grids.
Overall, these results highlight the superior spatial flexibility of the proposed XL-MA placement, which enables reliable signal coverage even under partial visibility conditions.

To evaluate the impact of NLoS components on system performance, Fig.~\ref{fig:Scenario2Rician} shows the expected weighted sum rate under MRC and MMSE for different antenna placement schemes with varying Rician factor $\kappa$. 
As $\kappa$ increases, the performance of all schemes degrades due to the diminishing contribution of NLoS paths and the resulting reduction in total channel power. For small Rician factors (i.e., rich scattering environments), the performance gap between the proposed MA and benchmark schemes is relatively small under both MRC and MMSE, as the enhanced NLoS power improves spatial diversity and reduces user channel correlation even in fixed-placement scenarios.
However, as $\kappa$ becomes large (i.e., LoS-dominant), the proposed XL-MA placement achieves significantly higher rates, especially under MMSE receiver. This is because fixed subarray layouts suffer from strong spatial correlation among users when the channel is dominated by LoS components, whereas the proposed XL-MA placement can adaptively position subarrays to decorrelate LoS steering vectors and exploit spatial non-stationarity.
These results highlight that the proposed XL-MA scheme is more effective in LoS-dominant near-field environments or  those exhibiting channel path sparsity in high-frequency bands.

\subsection{Performance Under Different User Distributions}

\begin{figure}[t]
    \centering
        \includegraphics[width=0.8\linewidth]{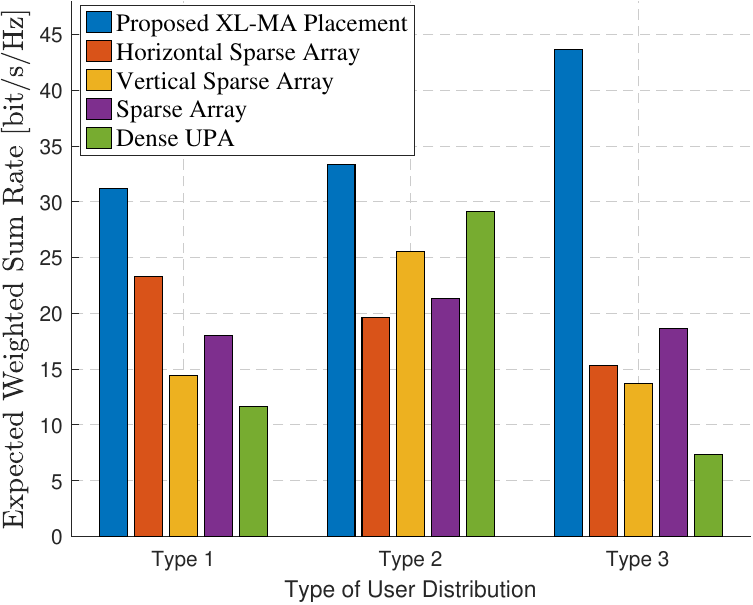} 
	 \vspace{-0.5em} 
    \caption{Performance under various user distributions.}
    \label{fig:Scenario3Performance}
    \vspace{-1.5em}
\end{figure}

In this subsection, we consider a general scenario with the 2D MA region and 3D coverage region under partial LoS visibility.
Therein, two obstacles are placed between the MA region and the coverage region as described in Section \ref{subsec:PartialLoS}.
Moreover, each subarray is configured as a UPA with $M_{\mathrm{H}} = 4$ and $M_{\mathrm{V}} = 4$.
We consider three types of user distributions, each with 12 hotspot grids of identical user density or probability, defined as follows:
(i) \textit{Type 1}: The hotspot grids are distributed along the $y$-axis, with their 3D indices specified by $\mathcal{K}_{1,1} = \{(1, k_y, 1) \mid k_y = \lfloor \frac{K_y - 1}{11} \cdot (k - 1) + 1 \rceil, k = 1, 2, \dots, 12\}$; (ii) \textit{Type 2}: The hotspot grids are distributed along the $z$-axis, with their indices specified by $\mathcal{K}_{2,1} = \{(1, \lceil \frac{K_y}{2} \rceil - 4, k_z) \mid  k_z = \lfloor 1 + \frac{((k-1) \bmod 7)(K_z - 1)}{5} \rceil, k = 1, 2, \dots, 6\} \cup \{(1, \lceil \frac{K_y}{2} \rceil + 4, k_z) \mid  k_z = \lfloor 1 + \frac{((k-1) \bmod 7)(K_z - 1)}{5} \rceil, k = 7, 8, \dots, 12\}$; (iii) \textit{Type 3}: The hotspot grids are distributed in the $yz$-plane with their indices specified by $\mathcal{K}_{3,1} = \{(2, k_y, K_z-2) \mid  k_y \in\{2,3,4,K_y-3,K_y-2,K_y-1\}\} \cup \{(2, k_y, K_z) \mid  k_y \in\{2,3,4,K_y-3,K_y-2,K_y-1\}\}$.

Fig. \ref{fig:Scenario3Performance} shows the expected weighted sum-rate under MRC with different user distribution, with $\bar{K} = 10$, $\kappa = 20$ dB, and $\zeta = 0$.
It is observed that the proposed XL-MA placement outperforms all FPA benchmarks across various user distributions, with particularly significant gains under 3D user distributions (i.e., Type 3) due to its flexibility in adapting to non-uniform user distribution.
In contrast, while one of FPA benchmarks achieves relatively good performance under a specific user distribution, all FPA schemes suffer from limited adaptability to varying user distributions, leading to significant performance degradation in practical scenarios with heterogeneous user distributions.

\section{Conclusion}\label{sec:Conclusion}

This paper investigated an uplink multiuser communication system employing multiple MA subarrays, where each subarray can be flexibly positioned within a large spatial region to dynamically adapt to  user distributions and/or statistical channel conditions.
In  particular, we developed a spatially non-stationary channel model that accounts for near-field characteristics, including large-scale path gains, AoAs, and LoS visibility. 
A closed-form approximation of the expected weighted sum rate under MRC was derived, demonstrating that optimized subarray placement can significantly boost user channel power gain and reduce channel correlation.
To solve the subarray placement optimization problem efficiently, we proposed a polynomial-time successive replacement algorithm. 
Simulation results validated the effectiveness of the proposed XL-MA placement strategy, achieving near-optimal performance and outperforming benchmark schemes based on FPAs under both MRC and MMSE combining.

\appendix
\subsection{Proof of Theorem \ref{theorem:MRCapprox}} \label{theorem:MRCapprox_proof}

By applying the approximation $\mathbb{E} \left\{ \log_2 \left( 1 + {X}/{Y} \right) \right\} \approx \log_2 \left( 1 + {\mathbb{E} \{X\}}/{\mathbb{E} \{Y\}} \right)$ from \cite{YANG2020Uplink}, we can approximate $R_k^{\text{MR}}$ in \eqref{Eq:RateMR} $  \tilde{R}_k^{\text{MR}} =   \log_2\left(1+ \tilde{\gamma}_k^{\text{MR}}\right)$,
with
{\small
\begin{align} \label{Eq:SINRMR1}	 
	& \tilde{\gamma}_k^{\text{MR}} = \nonumber\\
	&\frac{\bar{P}_k \mathbb{E}\{\left|\bm{\bar h}^{\sf H}_{k} (\bm {\Phi}^{\sf H}\bm {\Phi} \otimes \mathbf{I}_{M})\bm{\bar h}_{k}\right|^2\}}{\sum\limits_{\substack{i=1,\\i\neq k}}^K \bar{P}_i \rho_i\mathbb{E}\{\left|\bm{\bar h}^{\sf H}_{k} (\bm {\Phi}^{\sf H}\bm {\Phi} \otimes \mathbf{I}_{M})\bm{\bar h}_{i}\right|^2\} + \mathbb{E}\{\bm{\bar h}^{\sf H}_{k} (\bm {\Phi}^{\sf H}\bm {\Phi} \otimes \mathbf{I}_{M})\bm{\bar h}_{k}\}},
\end{align}}where $\mathbb{E}\{\alpha_i\} = \rho_i$. 
First, we derive the term $\mathbb{E} \{ \bm{\bar h}^{\sf H}_{k} (\bm {\Phi}^{\sf T}\bm {\Phi} \otimes \mathbf{I}_{M})\bm{\bar h}_{k} \}, \forall k$. 
It follows from  constraints \eqref{P0Cons2} and \eqref{P0Cons1} that 
$[\bm {\Phi}^{\sf H}\bm {\Phi}]_{jk} = \sum_{i = 1}^{N}[\bm {\Phi}]_{ij}[\bm {\Phi}]_{ik} = 0$, $\forall j, i, j \neq k$ and $[\bm {\Phi}^{\sf H}\bm {\Phi}]_{jj} = \sum_{i = 1}^{N}[\bm {\Phi}]_{ij} = [\bm{1}_N^{\sf T}\bm{\Phi}]_j, \forall j$, i.e.,
\begin{align} \label{Eq:proofA1}
    \bm {\Phi}^{\sf H}\bm {\Phi} = \text{diag}(\bm{1}_N^{\sf T}\bm{\Phi}).
\end{align}
By substituting \eqref{Eq:proofA1} into $\mathbb{E} \{ \bm{\bar h}^{\sf H}_{k} (\bm {\Phi}^{\sf T}\bm {\Phi} \otimes \mathbf{I}_{M})\bm{\bar h}_{k} \}$, it reduces to
\begin{align}\label{Eq:proofA3}
		\mathbb{E} \left\{ \bm{\bar h}^{\sf H}_{k} (\bm {\Phi}^{\sf T}\bm {\Phi} \otimes \mathbf{I}_{M})\bm{\bar h}_{k} \right\} = \sum_{\tilde{n} = 1}^{N_0} [\bm{1}_N^{\sf T}\bm{\Phi}]_{\tilde{n}} \mathbb{E} \{ \| \tilde{\bm{h}}_{k}(\mathbf{r}_{\tilde{n}} )\|^2 \}.
\end{align}
Then, we evaluate $\mathbb{E} \big\{ \| \tilde{\bm{h}}_{k}(\mathbf{r}_{\tilde{n}} )\|^2 \big\}$ in \eqref{Eq:proofA3}, which is given by
\begin{align} \label{Eq:proofA4}
  \mathbb{E} \big\{\| \tilde{\bm{h}}_{k}(\mathbf{r}_{\tilde{n}}) \|^2 \} 
  &\stackrel{(a)}{=} M \big(\xi_{\text{LoS},k}(\mathbf{r}_{\tilde{n}}) \beta_{\text{LoS},k}(\mathbf{r}_{\tilde{n}}) + \beta_{\text{NLoS},k}(\mathbf{r}_{\tilde{n}})\big) \nonumber\\
  &= M\beta_k(\mathbf{r}_{\tilde{n}}),
\end{align}
where $(a)$ holds due to the independence of the LoS and NLoS components, both of which have zero mean.
By substituting \eqref{Eq:proofA4} into \eqref{Eq:proofA3}, we obtain	
\begin{align} \label{Eq:eq:proofA10}
	\mathbb{E} \left\{ \bm{\bar h}^{\sf H}_{k} (\bm {\Phi}^{\sf T}\bm {\Phi} \otimes \mathbf{I}_{M})\bm{\bar h}_{k} \right\} =  M\bm{1}^{\sf T}_N\bm{\Phi}\bm{\bar{\beta}}_k.
\end{align}

Next, we derive the term $\mathbb{E}\{\left|\bm{\bar h}^{\sf H}_{k} (\bm {\Phi}^{\sf H}\bm {\Phi} \otimes \mathbf{I}_{M})\bm{\bar h}_{k}\right|^2\}$.
With \eqref{Eq:proofA1}, it reduces to
{\small
\begin{align}\label{Eq:proofA11}
    &\!\!\mathbb{E}\{\left|\bm{\bar h}^{\sf H}_{k} (\bm {\Phi}^{\sf H}\bm {\Phi} \otimes \mathbf{I}_{M})\bm{\bar h}_{k}\right|^2\} = 
	  \sum_{\tilde{n} = 1}^{N_0} [\bm{1}_N^{\sf T}\bm{\Phi}]_{\tilde{n}} \mathbb{E} \{ \| \tilde{\bm{h}}_{k}(\mathbf{r}_{\tilde{n}} )\|^4 \} \nonumber \\
    &\!\! + 2 \sum_{\tilde{n}=1}^{N_0} \sum_{\tilde{n}' > \tilde{n}}^{N_0}[\bm{1}_N^{\sf T}\bm{\Phi}]_{\tilde{n}}[\bm{1}_N^{\sf T}\bm{\Phi}]_{\tilde{n}'} \mathbb{E} \{ \| \tilde{\bm{h}}_{k}(\mathbf{r}_{\tilde{n}} )\|^2  \| \tilde{\bm{h}}_{k}(\mathbf{r}_{\tilde{n}'} )\|^2 \}.
\end{align}
}The term $\mathbb{E}\{ \| \tilde{\bm{h}}_{k}(\mathbf{r}_{\tilde{n}} )\|^2  \| \tilde{\bm{h}}_{k}(\mathbf{r}_{\tilde{n}'} )\|^2 \}$ in \eqref
{Eq:proofA11} is given by
\begin{align}\label{Eq:proofA12}
    \mathbb{E} \{ \| \tilde{\bm{h}}_{k}(\mathbf{r}_{\tilde{n}} )\|^2  \| \tilde{\bm{h}}_{k}(\mathbf{r}_{\tilde{n}'} )\|^2 \}
	& \stackrel{(b)}{=}  M^2\beta_k(\mathbf{r}_{\tilde{n}})\beta_k(\mathbf{r}_{\tilde{n}'}).
\end{align} 
where $(b)$ holds due to the independence of $\tilde{\bm{h}}_{k}(\mathbf{r}_{\tilde{n}} )$ and $\tilde{\bm{h}}_{k}(\mathbf{r}_{\tilde{n}'} )$ for $\tilde{n} \neq \tilde{n}'$, as well as \eqref{Eq:proofA4}.
Moreover, the term $\mathbb{E} \{ \| \tilde{\bm{h}}_{k}(\mathbf{r}_{\tilde{n}} )\|^4 \}$ in \eqref{Eq:proofA11} is computed by expanding it as
{\small
\begin{align} \label{Eq:proofA13}
 &\mathbb{E} \{ \| \tilde{\bm{h}}_{k}(\mathbf{r}_{\tilde{n}} )\|^4 \} = M\beta^2_{\text{NLoS},k}(\mathbf{r}_{\tilde{n}}) \nonumber\\
	& \hspace{2em}+ M^2 \beta^2_k(\mathbf{r}_{\tilde{n}}) + M2\xi_{\text{LoS}, k}(\mathbf{r}_{\tilde{n}})\beta_{\text{LoS},k}(\mathbf{r}_{\tilde{n}}) \beta_{\text{NLoS},k}(\mathbf{r}_{\tilde{n}}).
\end{align} 
}
By substituting \eqref{Eq:proofA12} and \eqref{Eq:proofA13} into \eqref{Eq:proofA11}, $\mathbb{E}\{\left|\bm{\bar h}^{\sf H}_{k} (\bm {\Phi}^{\sf H}\bm {\Phi} \otimes \mathbf{I}_{M})\bm{\bar h}_{k}\right|^2\}$ can be expressed as
\begin{align}\label{Eq:proofA21}
     M^2 \left( \bm{1}_N^{\sf T}\bm{\Phi} \bm{\bar \beta}_{k} \right)^2 + M \bm{1}_N^{\sf T}\bm{\Phi}(\bm{\bar{\beta}}_k^{\circ 2} \circ\bm{\bar{f}}_k),
\end{align}
where the auxiliary vector $\bm{\bar{f}}_k \in \mathbb{R}^{N_0}$ is defined as
\begin{align} \label{Eq:fdefine}
[\bm{\bar{f}}_k]_{\tilde{n}} &=  \frac{M(2[{\bar{\bm \kappa}_{k}}]_{\tilde{n}}\xi_{\text{LoS}, k}(\mathbf{r}_{\tilde{n}})+ 1)}{([{\bar{\bm \kappa}_{k}}]_{\tilde{n}}\xi_{\text{LoS}, k}(\mathbf{r}_{\tilde{n}})+1)^2}, \forall \tilde{n},
\end{align}
with 
$[{\bar{\bm \kappa}_{k}}]_{\tilde{n}} = \frac{\beta_{\text{LoS},k}(\mathbf{r}_{\tilde{n}})}{\beta_{\text{NLoS},k}(\mathbf{r}_{\tilde{n}})}$, $1\leq \tilde{n}\leq N_0$.

Finally, we derive the term $\mathbb{E}\{\left|\bm{\bar h}^{\sf H}_{k} (\bm {\Phi}^{\sf H}\bm {\Phi} \otimes \mathbf{I}_{M})\bm{\bar h}_{i}\right|^2\}$ for $i \neq k$, which is derived by substituting \eqref{Eq:proofA1} as
{\small 
\begin{align}\label{Eq:proofA22}
    \mathbb{E}\{|\bm{\bar h}^{\sf H}_{k} (\bm {\Phi}^{\sf H}\bm {\Phi} \otimes \mathbf{I}_{M})\bm{\bar h}_{i}|^2\}& \stackrel{(c)}{=} \sum_{\tilde{n} = 1}^{N_0} [\bm{1}_N^{\sf T}\bm{\Phi}]_{\tilde{n}} \mathbb{E} \{ | \tilde{\bm{h}}_{k}(\mathbf{r}_{\tilde{n}})^{\sf H} \tilde{\bm{h}}_{i}(\mathbf{r}_{\tilde{n}})|^2\},
\end{align}}where $(c)$ holds due to the independence of $\tilde{\bm{h}}_{k}(\mathbf{r}_{\tilde{n}})$ and $\tilde{\bm{h}}_{i}(\mathbf{r}_{\tilde{n}})$ for $k \neq i$, as well as their zero means,
resulting in  $\mathbb{E} \{ \tilde{\bm{h}}_{k}(\mathbf{r}_{\tilde{n}'})^{\sf H} \tilde{\bm{h}}_{i}(\mathbf{r}_{\tilde{n}'})\tilde{\bm{h}}_{k}(\mathbf{r}_{\tilde{n}'})^{\sf H} \tilde{\bm{h}}_{i}(\mathbf{r}_{\tilde{n}'})\} = 0$, $\forall \tilde{n}', \tilde{n}$, $\tilde{n}' \neq \tilde{n}$.
By expanding the squared norm of $\tilde{\bm{h}}_{k}(\mathbf{r}_{\tilde{n}})^{\sf H} \tilde{\bm{h}}_{i}(\mathbf{r}_{\tilde{n}})$ in \eqref{Eq:proofA22} and removing the terms with zero mean, we have
{\small
\begin{align}\label{Eq:proofA23}
	& \mathbb{E} \{| \tilde{\bm{h}}_{k}(\mathbf{r}_{\tilde{n}})^{\sf H} \tilde{\bm{h}}_{i}(\mathbf{r}_{\tilde{n}})|^2\} 
	 \!=\! \xi_{\text{LoS}, k}(\mathbf{r}_{\tilde{n}}) \xi_{\text{LoS}, i}(\mathbf{r}_{\tilde{n}})\beta_{\text{LoS},k}(\mathbf{r}_{\tilde{n}})\beta_{\text{LoS},i}(\mathbf{r}_{\tilde{n}})[\bm {\bar \varphi}_{k,i}]_{\tilde{n}} \nonumber\\
	& +  M \beta_{\text{NLoS},k}(\mathbf{r}_{\tilde{n}}) \beta_{\text{NLoS},i}(\mathbf{r}_{\tilde{n}}) \! +\! M\xi_{\text{LoS}, k}(\mathbf{r}_{\tilde{n}}) \beta_{\text{LoS},k}(\mathbf{r}_{\tilde{n}}) \beta_{\text{NLoS},i}(\mathbf{r}_{\tilde{n}}) \nonumber\\
	&+M\xi_{\text{LoS}, i}(\mathbf{r}_{\tilde{n}}) \beta_{\text{NLoS},k}(\mathbf{r}_{\tilde{n}}) \beta_{\text{LoS},i}(\mathbf{r}_{\tilde{n}}) 
\end{align}
}By substituting \eqref{Eq:proofA23} into \eqref{Eq:proofA22}, the term $\mathbb{E}\{\left|\bm{\bar h}^{\sf H}_{k} (\bm {\Phi}^{\sf H}\bm {\Phi} \otimes \mathbf{I}_{M})\bm{\bar h}_{i}\right|^2\}$  can be expressed as
\begin{align}\label{Eq:proofA29}
	\bm{1}_N^{\sf T}\bm{\Phi} (\bm{\bar \beta}_{k}\circ\bm{\bar{\beta}}_i \circ (\bm{\bar{\varphi}}_{k,i}\circ \bm{\bar{g}}_{k,i} + \bm{\bar{q}}_{k,i})),
\end{align}
where the auxiliary vectors $\bm{\bar{g}}_{k,i} \in \mathbb{R}^{N_0}$ and $\bm{\bar{q}}_{k,i} \in \mathbb{R}^{N_0}$ are respectively defined as
\begin{align} 
&\!\!\![\bm{\bar{g}}_{k,i}]_{\tilde{n}} =  \frac{[{\bar{\bm \kappa}_{k}}]_{\tilde{n}}[{\bar{\bm \kappa}_{i}}]_{\tilde{n}}\xi_{\text{LoS}, k}(\mathbf{r}_{\tilde{n}})\xi_{\text{LoS}, i}(\mathbf{r}_{\tilde{n}})}{([{\bar{\bm \kappa}_{k}}]_{\tilde{n}}\xi_{\text{LoS}, k}(\mathbf{r}_{\tilde{n}})+1)([{\bar{\bm \kappa}_{i}}]_{\tilde{n}}\xi_{\text{LoS}, i}(\mathbf{r}_{\tilde{n}})+1)}, \forall \tilde{n}, \label{Eq:gdefine} \\
&\!\!\!	[\bm{\bar{q}}_{k,i}]_{\tilde{n}} =  \frac{ (1+[{\bar{\bm \kappa}_{k}}]_{\tilde{n}}\xi_{\text{LoS}, k}(\mathbf{r}_{\tilde{n}})+[{\bar{\bm \kappa}_{i}}]_{\tilde{n}}\xi_{\text{LoS}, i}(\mathbf{r}_{\tilde{n}}))  M}{([{\bar{\bm \kappa}_{k}}]_{\tilde{n}}\xi_{\text{LoS}, k}(\mathbf{r}_{\tilde{n}})+1)([{\bar{\bm \kappa}_{i}}]_{\tilde{n}}\xi_{\text{LoS}, i}(\mathbf{r}_{\tilde{n}})+1)}, \forall \tilde{n}. \label{Eq:qdefine}
\end{align}

In summary, by substituting \eqref{Eq:eq:proofA10}, \eqref{Eq:proofA21}, and \eqref{Eq:proofA29} into \eqref{Eq:SINRMR1}, we obtain $\tilde{\gamma}_k^{\text{MR}}$ in \eqref{Neq:sinrMRapprox}.
This thus completes the proof.
\hfill $\square$

\bibliographystyle{IEEEtran}
\bibliography{Ref} 

\begin{thebibliography}{10}
\providecommand{\url}[1]{#1}
\csname url@samestyle\endcsname
\providecommand{\newblock}{\relax}
\providecommand{\bibinfo}[2]{#2}
\providecommand{\BIBentrySTDinterwordspacing}{\spaceskip=0pt\relax}
\providecommand{\BIBentryALTinterwordstretchfactor}{4}
\providecommand{\BIBentryALTinterwordspacing}{\spaceskip=\fontdimen2\font plus
\BIBentryALTinterwordstretchfactor\fontdimen3\font minus
  \fontdimen4\font\relax}
\providecommand{\BIBforeignlanguage}[2]{{%
\expandafter\ifx\csname l@#1\endcsname\relax
\typeout{** WARNING: IEEEtran.bst: No hyphenation pattern has been}%
\typeout{** loaded for the language `#1'. Using the pattern for}%
\typeout{** the default language instead.}%
\else
\language=\csname l@#1\endcsname
\fi
#2}}
\providecommand{\BIBdecl}{\relax}
\BIBdecl

\bibitem{LU2024Tutorial}
H.~Lu, Y.~Zeng \emph{et~al.}, ``A tutorial on near-field {XL}-{MIMO}
  communications towards 6{G},'' \emph{IEEE Commun. Surv. Tutorials}, vol.~26,
  no.~4, pp. 2213--2257, Apr. 2024.

\bibitem{WANG2024Tutorial}
Z.~Wang, J.~Zhang \emph{et~al.}, ``A tutorial on extremely large-scale {MIMO}
  for 6g: Fundamentals, signal processing, and applications,'' \emph{IEEE
  Commun. Surv. Tutor.}, vol.~26, no.~3, pp. 1560--1605, Jul. 2024.

\bibitem{sanayei2004antenna}
S.~Sanayei and A.~Nosratinia, ``Antenna selection in {MIMO} systems,''
  \emph{IEEE Commun. Mag.}, vol.~42, no.~10, pp. 68--73, Oct. 2004.

\bibitem{li2025sparse}
X.~Li, H.~Min \emph{et~al.}, ``Sparse {MIMO} for {ISAC}: New opportunities and
  challenges,'' \emph{IEEE Wireless Commun., early access}, 2025.

\bibitem{Wang2023SparseArrays}
H.~Wang and Y.~Zeng, ``Can sparse arrays outperform collocated arrays for
  future wireless communications?'' in \emph{Proc. IEEE GLOBECOM Workshops (GC
  Wkshps)}, 2023, pp. 667--672.

\bibitem{ZHU2024Movable}
L.~Zhu, W.~Ma, and R.~Zhang, ``Movable antennas for wireless communication:
  Opportunities and challenges,'' \emph{IEEE Commun. Mag.}, vol.~62, no.~6, pp.
  114--120, Jun. 2024.

\bibitem{ZHU2024Modeling}
------, ``Modeling and performance analysis for movable antenna enabled
  wireless communications,'' \emph{IEEE Trans. Wireless Commun.}, vol.~23,
  no.~6, pp. 6234--6250, Jun. 2024.

\bibitem{zhu2025tutorial}
L.~Zhu, W.~Ma \emph{et~al.}, ``A tutorial on movable antennas for wireless
  networks,'' \emph{IEEE Commun. Surv. Tutor., early access}, 2025.

\bibitem{wong2022bruce}
K.-K. Wong, K.-F. Tong \emph{et~al.}, ``Bruce lee-inspired fluid antenna
  system: Six research topics and the potentials for 6{G},'' \emph{Front.
  Comms. Net.,}, vol.~3, no. 853416, pp. 1--31, Mar, 2022.

\bibitem{MEI2024MovableAntennaa}
W.~Mei, X.~Wei \emph{et~al.}, ``Movable-antenna position optimization: A
  graph-based approach,'' \emph{IEEE Wirel. Commun. Lett.}, vol.~13, no.~7, pp.
  1853--1857, Jul. 2024.

\bibitem{zhu2023movable}
L.~Zhu, W.~Ma, and R.~Zhang, ``Movable-antenna array enhanced beamforming:
  Achieving full array gain with null steering,'' \emph{IEEE Commun. Lett.},
  vol.~27, no.~12, pp. 3340--3344, Dec. 2023.

\bibitem{ma2024multibeam}
W.~Ma, L.~Zhu, and R.~Zhang, ``Multi-beam forming with movable-antenna array,''
  \emph{IEEE Commun. Lett.}, vol.~28, no.~3, pp. 697--701, Mar. 2024.

\bibitem{MA2024MIMO}
------, ``{MIMO} capacity characterization for movable antenna systems,''
  \emph{IEEE Trans. Wireless Commun.}, vol.~23, no.~4, pp. 3392--3407, Apr.
  2024.

\bibitem{ZHU2024MovableAntenna}
L.~Zhu, W.~Ma, B.~Ning, and R.~Zhang, ``Movable-antenna enhanced multiuser
  communication via antenna position optimization,'' \emph{IEEE Trans. Wireless
  Commun.}, vol.~23, no.~7, pp. 7214--7229, Jul. 2024.

\bibitem{Xiao2024Multiuser}
Z.~Xiao, X.~Pi \emph{et~al.}, ``Multiuser communications with movable-antenna
  base station: Joint antenna positioning, receive combining, and power
  control,'' \emph{IEEE Trans. Wireless Commun.}, vol.~23, no.~12, pp.
  19\,744--19\,759, Dec. 2024.

\bibitem{Tang2025Secure}
J.~Tang, C.~Pan \emph{et~al.}, ``Secure {MIMO} communication relying on movable
  antennas,'' \emph{IEEE Trans. Commun.}, vol.~73, no.~4, pp. 2159--2175, Apr.
  2025.

\bibitem{qin2024antenna}
H.~Qin, L.~Zhu, and R.~Zhang, ``Antenna positioning and beamforming design for
  fluid antenna-assisted multi-user downlink communications,'' \emph{IEEE
  Wireless Commun. Lett.}, vol.~13, no.~4, pp. 1073--1077, Apr. 2024.

\bibitem{Hu2025Movable}
G.~Hu, Q.~Wu \emph{et~al.}, ``Movable antennas-enabled two-user multicasting:
  Do we really need alternating optimization for minimum rate maximization?''
  \emph{IEEE Trans. Veh. Technol.}, vol.~74, no.~3, pp. 5135--5140, Mar. 2025.

\bibitem{hu2024fluid}
------, ``Fluid antennas-enabled multiuser uplink: A low-complexity gradient
  descent for total transmit power minimization,'' \emph{IEEE Commun. Lett.},
  vol.~28, no.~3, pp. 602--606, Mar. 2024.

\bibitem{LU2024Group}
H.~Lu, Y.~Zeng, S.~Jin, and R.~Zhang, ``Group movable antenna with flexible
  sparsity: Joint array position and sparsity optimization,'' \emph{IEEE
  Wireless Commun. Lett.}, vol.~13, no.~12, pp. 3573--3577, Dec. 2024.

\bibitem{Ma2023Compressed}
W.~Ma, L.~Zhu, and R.~Zhang, ``Compressed sensing based channel estimation for
  movable antenna communications,'' \emph{IEEE Commun. Lett.}, vol.~27, no.~10,
  pp. 2747--2751, Oct. 2023.

\bibitem{Xu2024Channel}
H.~Xu, G.~Zhou \emph{et~al.}, ``Channel estimation for {FAS}-assisted multiuser
  mmwave systems,'' \emph{IEEE Commun. Lett.}, vol.~28, no.~3, pp. 632--636,
  Mar. 2024.

\bibitem{ye2023fluid}
Y.~Ye, L.~You, J.~Wang, H.~Xu, K.-K. Wong, and X.~Gao, ``Fluid antenna-assisted
  {MIMO} transmission exploiting statistical {CSI},'' \emph{IEEE Commun.
  Lett.}, vol.~28, no.~1, pp. 223--227, Jan. 2023.

\bibitem{hu2024two}
G.~Hu, Q.~Wu \emph{et~al.}, ``Two-timescale design for movable antenna
  array-enabled multiuser uplink communications,'' \emph{IEEE Trans. Veh.
  Technol.}, vol.~73, no.~4, pp. 3922--3936, Apr. 2024.

\bibitem{zheng2024two}
Z.~Zheng, Q.~Wu, W.~Chen, and G.~Hu, ``Two-timescale design for movable
  antennas enabled-multiuser {MIMO} systems,'' \emph{arXiv preprint
  arXiv:2410.05912}, 2024.

\bibitem{zhu2025movable}
L.~Zhu, W.~Ma, Z.~Xiao, and R.~Zhang, ``Movable antenna enabled near-field
  communications: Channel modeling and performance optimization,'' \emph{IEEE
  Trans. Commun., early access}, 2025.

\bibitem{yan2025movable}
G.~Yan, L.~Zhu, and R.~Zhang, ``Movable antenna aided multiuser communications:
  Antenna position optimization based on statistical channel information,''
  \emph{arXiv preprint arXiv:2502.20856}, 2025.

\bibitem{shao20246d}
X.~Shao, Q.~Jiang, and R.~Zhang, ``6{D} movable antenna based on user
  distribution: Modeling and optimization,'' \emph{IEEE Trans. Wireless
  Commun.}, vol.~23, no.~6, pp. 6240--6255, Jun. 2024.

\bibitem{SHAO20256Da}
X.~Shao, R.~Zhang \emph{et~al.}, ``6{D} movable antenna enhanced wireless
  network via discrete position and rotation optimization,'' \emph{IEEE J.
  Select. Areas Commun.}, vol.~43, no.~3, pp. 674--687, Mar. 2025.

\bibitem{SHAO20256DMA}
X.~Shao and R.~Zhang, ``6{DMA} enhanced wireless network with flexible antenna
  position and rotation: Opportunities and challenges,'' \emph{IEEE Commun.
  Mag.}, vol.~63, no.~4, pp. 121--128, Apr. 2025.

\bibitem{Ding2025Flexible}
Z.~Ding, R.~Schober, and H.~Vincent~Poor, ``Flexible-antenna systems: A
  pinching-antenna perspective,'' \emph{IEEE Trans. Commun., early access},
  2025.

\bibitem{liu2025pinching}
Y.~Liu, Z.~Wang \emph{et~al.}, ``Pinching antenna systems ({PASS}):
  Architecture designs, opportunities, and outlook,'' \emph{arXiv preprint
  arXiv:2501.18409}, 2025.

\bibitem{ouyang2025array}
C.~Ouyang, Z.~Wang \emph{et~al.}, ``Array gain for pinching-antenna systems
  ({PASS}),'' \emph{arXiv preprint arXiv:2501.05657}, 2025.

\bibitem{fu2024multi}
M.~Fu, L.~Zhu, and R.~Zhang, ``Multi-{IRS} enhanced wireless coverage:
  Deployment optimization based on large-scale channel knowledge,'' \emph{arXiv
  preprint arXiv:2410.20042}, 2024.

\bibitem{marzetta2016fundamentals}
T.~L. Marzetta, E.~G. Larsson, H.~Yang, and H.~Q. Ngo, \emph{Fundamentals of
  massive {MIMO}}.\hskip 1em plus 0.5em minus 0.4em\relax Cambridge University
  Press, 2016.

\bibitem{Carvalho2020NonStationarities}
E.~D. Carvalho, A.~Ali \emph{et~al.}, ``Non-stationarities in extra-large-scale
  massive {MIMO},'' \emph{IEEE Wireless Commun.}, vol.~27, no.~4, pp. 74--80,
  Aug. 2020.

\bibitem{bjornson2017massive}
E.~Bj{\"o}rnson, J.~Hoydis, L.~Sanguinetti \emph{et~al.}, ``Massive {MIMO}
  networks: Spectral, energy, and hardware efficiency,'' \emph{Found. Trends
  Signal Process.}, vol.~11, no. 3-4, pp. 154--655, 2017.

\bibitem{YANG2020Uplink}
X.~Yang, F.~Cao, M.~Matthaiou, and S.~Jin, ``On the uplink transmission of
  extra-large scale massive {MIMO} systems,'' \emph{IEEE Trans. Veh. Technol.},
  vol.~69, no.~12, pp. 15\,229--15\,243, Dec. 2020.

\end{thebibliography}
\end{document}